\documentclass[12pt]{article}
\usepackage{latexsym, amssymb}
\def\beq{\begin{equation}}
\def\eeq{\end{equation}}
\def\beqs{\begin{equation*}}
\def\eeqs{\end{equation*}}
\def\bea{\begin{eqnarray}}
\def\eea{\end{eqnarray}}
\def\leq{\lefteqn}
\def\bay{\begin{array}}
\def\eay{\end{array}}
\def\beas{\begin{eqnarray*}}
\def\eeas{\end{eqnarray*}}
\def\bq{\begin{quote}}
\def\eq{\end{quote}}

\def\p{\partial}
\def\bone{{\mbox{\boldmath$1$}}}
\def\a{\alpha}
\def\ba{{\mbox{\boldmath$\alpha$}}}
\def\bsa{{\mbox{\boldmath$\alpha^*$}}}
\def\sa{\alpha^*}
\def\ad{a^\dgr}

\def\b{\beta}
\def\bb{{\mbox{\boldmath$\beta$}}}
\def\bbs{{\mbox{\boldmath$\beta^*$}}}
\def\sb{\beta^*}

\def\c{\gamma}

\def\bc{{\mbox{\boldmath$\gamma$}}}
\def\bct{{\mbox{\scriptsize\boldmath$\gamma$}}}
\def\bcs{{\mbox{\boldmath$\gamma^*$}}}
\def\bcst{{\mbox{\scriptsize\boldmath$\gamma^*$}}}

\def\sc{\gamma^*}
\def\cs{\gamma^*}

\def\d{\delta}

\def\e{\varepsilon}

\def\l{\lambda}

\def\th{\theta}

\def\s{\sigma}

\def\z{\zeta}
\def\bz{{\mbox{\boldmath$\zeta$}}}
\def\sz{\zeta^*}
\def\zs{\zeta^*}

\def\x{\xi}
\def\bx{{\mbox{\boldmath$\xi$}}}
\def\bxt{{\mbox{\scriptsize\boldmath$\xi$}}}
\def\bxs{{\mbox{\boldmath$\xi^*$}}}
\def\bxst{{\mbox{\scriptsize\boldmath$\xi^*$}}}
\def\xs{\xi^*}
\def\sx{\xi^*}
\def\nn{\nonumber}

\def\dgr{\dagger}

\def\t{\tilde}

\def\Tr{{\rm Tr}}
\def\t2{{\textstyle{2}}}
\def\half{\frac{1}{2}}
\def\thalf{\textstyle{ {1 \over 2} } }
\def\shalf{\scriptstyle{\frac{1}{2}}}

\def\sqr2{\sqrt{2}}

\def\G{Grassmann }

\def\eis{eigenstate }
\def\eiss{eigenstates }
\def\eissp{eigenstates}

\def\eivs{eigenvalues }
\def\cost{coherent state }
\def\costs{coherent states }
\def\co_st{coherent-state}
\def\disp{displacement }
\def\char{characteristic }
\def\op{operator }
\def\opp{operator}
\def\ops{operators }
\def\opsp{operators}
\def\dop{density operator }

\def\dopsp{density operators}
\def\ani{annihilation }
\def\cre{creation }
\def\and{\quad {\hbox \mathrm{and}} \quad}

\def\gappeq{\mathrel{\rlap {\raise.5ex\hbox{$>$}}
{\lower.5ex\hbox{$\sim$}}}}
\def\lappeq{\mathrel{\rlap{\raise.5ex\hbox{$<$}}
{\lower.5ex\hbox{$\sim$}}}}

\begin{document}
\pagestyle{empty}
\begin{flushright}
NMCPP/97-15\\
physics/9808029\\
\end{flushright}
\begin{center}
{\bf\Large Density Operators for Fermions}\\
\vspace*{1.0cm}
{\bf Kevin E.~Cahill\footnote{kevin@kevin.phys.unm.edu
 http://kevin.phys.unm.edu/$\tilde{\ }$kevin/}}\\
\vspace{.2cm}
New Mexico Center for Particle Physics\\
University of New Mexico\\
Albuquerque, NM 87131-1156\\
\vspace{.50cm}
{\bf Roy J.~Glauber\footnote{glauber@physics.harvard.edu}}\\
\vspace{.2cm}
Lyman Laboratory of Physics\\
Harvard University\\
Cambridge, MA 02138\\
\vspace*{1.0cm}
{\bf Abstract}
\end{center}
\begin{quote}
The mathematical methods that have been
used to analyze the statistical properties
of boson fields, and in particular the 
coherence of photons in quantum optics, 
have their counterparts for Fermi fields.
The coherent states, the displacement operators,
the P-representation, and the other operator
expansions all possess surprisingly close 
fermionic analogues. 
These methods for describing the
statistical properties of fermions 
are based upon a practical calculus of anti-commuting variables. 
They are used to calculate correlation functions
and counting distributions 
for general systems of fermions.
\end{quote}
\vfill
\begin{flushleft}
\today\\
\end{flushleft}
\vfill\eject
\centerline{}
\vfill\eject

\setcounter{equation}{0}
\setcounter{page}{1}
\pagestyle{plain}

\section{Introduction}
The Pauli exclusion principle 
plays an essential role in
describing the behavior of the particles,
both simple and complex,
that we now call fermions.
It is known to play a key role in
determining the structure of the most fundamental
elements of matter.
These are systems like atoms,
in which the phase-space density of fermions,
electrons in this case, is quite high.
But when fermionic atoms move freely in space
or even when they are trapped electromagnetically,
their phase-space density is usually so low that the effects
of the exclusion principle remain completely hidden.
A number of recent developments, however,
point to the possibility of achieving much higher densities
of fermionic atoms both in electromagnetic traps
and in free space.
\par
The various methods of optical cooling that
have been developed for atomic beams
work as well for fermions as they do for bosons
and produce beams with temperatures of the order of $100~\mu ^\circ$K.
Cooling fermions evaporatively to still lower temperatures
poses a problem that requires a less direct solution.
Evaporative cooling becomes inefficient
for fermions since the exclusion principle tends
to suppress collisions of identical atoms.
It may be implemented nonetheless by sympathetic means~\cite{symcool},
\emph{e.g.,} by cooling bosonic atoms at the same time,
so that energy exchange still takes place freely.
It seems possible thus that the realization of degenerate Fermi gases
may become an important byproduct of Bose-Einstein condensation.
\par
The detection methods that will be used
in measurements on beams of cold fermionic atoms
will be essentially the same as those
now used on bosonic atoms cooled by optical
or evaporative means.
The measurements on bosons can be most conveniently
described, in fact, by mathematical methods that were
introduced in the context of quantum optics~\cite{rjg2,GDW}.
\par
Much of the work in quantum optics, we may recall,
is couched in the language of coherent states,
which are eigenstates of the photon annihilation operators.
They contain an intrinsically indefinite number
of quanta but can nonetheless be used as a basis
for describing all states of the electromagnetic field.
While pure coherent states are not physically attainable
in bosonic systems with fixed numbers of particles,
it likewise remains useful to describe boson fields
in terms of suitably weighted superpositions and mixtures
of coherent states.
The weight functions associated with these combinations
may be regarded as quasi-probability densities in the spaces of
coherent-state amplitudes.
The function $ P $ in the coherent-state representation
of the density operator~\cite{rjg2,GDW} plays this role;
other quasi-probability densities including
the Wigner function~\cite{GDW,kcrg1}
and the $ Q $ function~\cite{GDW,kcrg1} 
play similarly convenient roles
in representing the density operator.
\par
In the case of fermion fields,
the vacuum state is the only physically realizable eigenstate
of the annihilation operators.
It is possible, however, to define such eigenstates
in a formal way
and to put them to many of the same analytical uses as are made
of the bosonic coherent states.
Since fermion field variables anti-commute,
their eigenvalues must, as noted by Schwinger~\cite{JS4},
be anti-commuting numbers.
Such numbers are Grassmann variables.
They can be handled by means of the simple rules
of Grassmann algebra~\cite{b}, which we include here
so that the calculations may be self-contained.
\par
Within this context we formulate ways of expressing
and evaluating a broad range of the correlation functions
that are measured in experiments involving the counting of fermions.
Central to this task is the expression of the quantum-mechanical
density operator in terms of Grassmann variables. 
We develop a number of ways of doing that in general terms
and present a detailed discussion of the density operators
for chaotically excited fields.
Included among the latter is a particularly useful
Gaussian representation of the grand-canonical density operator
for fermion fields.
Having evaluated the statistically averaged correlation
functions, we apply them to fermion counting experiments
and illustrate their use in determining the counting distributions.
\par
We find throughout this work that
notwithstanding great mathematical differences,
many close parallels can be established between
the expressions evaluated for fermion fields
and the more familiar ones for boson fields.
In particular, for example,
we can construct a family of quasi-probability
densities, as functions of the Grassmann variables,
with properties parallel to those of the entire
family of quasi-probability densities for bosons,
including the $P$, $Q$, and Wigner functions.
We can then evaluate the mean values of ordered products
of fermion creation and annihilation operators 
by performing integrations over the Grassmann variables
while using the appropriate quasi-probability density
as a weight function.
In both cases, we trade an inhomogeneous commutation
relation and an ordering rule for a homogeneous commutation
relation and a quasi-probability density.
For boson fields the integrations
are taken over commuting variables,
which may be treated as if they were classical variables.
For fermions, on the other hand,
the integrations are over anti-commuting variables
that have no classical analogues.
The weight functions for these integrations
are nevertheless in one-to-one correspondence
with the quasi-probability densities for bosons,
so it seems appropriate to give them similar names.
We have followed that convention for several other
parallels as well.

\section{Notation}
Let us consider a system of fermions 
which may be described by
the creation $ a_n^\dgr $ and annihilation $ a_m $ operators
which satisfy the familiar but ever mysterious relations
\bea
\{ a_n , a_m^\dgr \} & = & \d_{nm} \\
\{ a_n , a_m  \} & = & 0 \\
\{ a_n^\dgr ,  a_m^\dgr \} & = & 0 \\
a_n | 0 \rangle & = & 0 .
\label {acrels}
\eea
in which $ | 0 \rangle $ is the vacuum state. 
\par
We shall use lower-case Greek letters 
to denote Grassmann variables.
These anti-commuting numbers
$ \c_n $ and their complex conjugates
$ \sc_n $ satisfy the convenient relations
\bea
\{ \c_n , \c_m  \} & = & 0 \\
\{ \sc_n , \c_m  \} & = & 0 \\
\{ \sc_n , \sc_m  \} & = & 0 .
\label {grels}
\eea
We shall also assume that Grassmann variables 
anti-commute with fermionic operators 
\beq
\{ \c_n , a_m \} = 0 
\label {gopa}
\eeq
and commute with bosonic operators.
And we make the arbitrary choice
that hermitian conjugation reverses
the order of all fermionic quantities,
both the operators and the Grassmann numbers.
Thus for instance
we have
\beq
\left( a_1 \b_2 \ad_3 \sc_4 \right)^\dgr
= \c_4 a_3 \sb_2 \ad_1 .
\label {exa}
\eeq

\section{Coherent States for Fermions}
\subsection{Displacement Operators}
For any set $ \bc = \{ \c_i \} $ of Grassmann variables,
let us define the unitary displacement \op
$ D ( \bc ) $ as the exponential
\beq
D ( \bc ) = \exp 
\left( \sum_i \left( \ad_i \c_i - \sc_i a_i \right) \right) .
\label {D}
\eeq
One of the useful properties of Grassmann numbers
is that when, as in the preceding definition,
they multiply fermionic \ani or \cre \opsp,
their anti-commutativity cancels that 
of the \opsp.  Thus the \ops $ \ad_i \c_i $
and $ \sc_j a_j $ simply commute for $ i \ne j$\@.
So we may rewrite the displacement \op as the product
\bea
D ( \bc ) & = & \prod_i \exp \left( \ad_i \c_i - \sc_i a_i \right) 
\label {defD}\\
& = & \prod_i \left[ 1 + \ad_i \c_i - \sc_i a_i 
+ \left( \ad_i a_i - \thalf \right) \sc_i \c_i \right] .
\label {Dp}
\eea
By the same token,
the \ani \op $ a_n $ commutes with all the
\ops $ \ad_i \c_i $ and $ \sc_j a_j $ when $ n \ne i $,
and so we may compute the displaced \ani \op 
by ignoring all modes but the $n$th:
\bea
D^\dgr( \bc ) a_n D ( \bc )
& = & \prod_i \exp \left( \sc_i a_i - \ad_i \c_i \right)
\, a_n \, \prod_j \exp \left( \ad_j \c_j - \sc_j a_j \right) \nn\\
& = & \exp \left( \sc_n a_n - \ad_n \c_n \right)
\, a_n \, \exp \left( \ad_n \c_n - \sc_n a_n \right) \nn\\
& = & \left( 1 - \ad_n \c_n - \thalf \sc_n a_n \ad_n \c_n \right)
\, a_n \, \left( 1 + \ad_n \c_n - \thalf \ad_n \c_n \sc_n a_n \right) \nn\\
& = & \left( 1 - \ad_n \c_n - \thalf \sc_n \c_n \right)
\, a_n \, \left( 1 + \ad_n \c_n + \thalf \sc_n \c_n \right) \nn\\
& = & a_n - \ad_n \c_n a_n + a_n \ad_n \c_n 
= a_n + \c_n .
\label {DaD}
\eea
Similarly
\beq
D^\dgr( \bc ) \ad_n D ( \bc ) = \ad_n + \sc_n .
\label {DadD}
\eeq
\par
We may use the Baker-Hausdorff identity
\beq
e^{A + B} = e^A \, e^B \, e^{-{ \shalf } [ A , B ]} ,
\label {BH}
\eeq
which holds whenever the commutator $ [ A , B ] $
commutes with both $ A $ and $ B $,
to write the displacement operator $ D( \ba ) $ 
in forms that are normally ordered
\bea
\exp \left( \sum_i \ad_i \c_i \right) \,
\exp \left( - \sum_i \sc_i a_i \right) 
\! \! \! & = & \! \! \!
\exp
\left( \sum_i \left( \ad_i \c_i - \sc_i a_i \right) \right)
e ^ { \shalf \bcst\cdot\bct } \nn\\
D_N( \bc ) \! \! \! & = & \! \! \! D ( \bc ) 
\, \exp \left( \thalf \sum_i \sc_i \c_i \right) 
\label {DN}
\eea
and anti-normally ordered
\bea
\exp \left( - \sum_i \sc_i a_i \right) 
\exp \left( \sum_i \ad_i \c_i \right) \! \! \! & = & \! \! \! 
\exp
\left( \sum_i \left( \ad_i \c_i - \sc_i a_i \right) \right) 
e ^ { - \shalf \bcst\cdot\bct } \nn\\
D_A( \bc ) \! \! \! & = & \! \! \! D ( \bc ) \, 
\exp \left( - \thalf \sum_i \sc_i \c_i \right) ,
\label {DA}
\eea
in which we have employed the concise notation
\beq
\bcs \cdot \bc \equiv \sum_i \cs_i \c_i ,
\label {dotnot}
\eeq
an abbreviation
which we shall use occasionally but not exclusively.
The identity (\ref{BH}) also allows one to show that
the displacement operators form a ray representation
of the additive group of Grassmann numbers,
\beq
D( \ba ) \, D( \bb ) =
D ( \ba + \bb ) \exp \left[ \thalf \sum_i
\left( \sb_i \a_i - \sa_i \b_i \right) \right] .
\label {DD}
\eeq
 
\subsection{Coherent States}
For any set $ \bc = \{ \c_i \} $ of Grassmann numbers,
we define the \emph{normalized\/} coherent state $ | \bc \rangle $ as
the displaced vacuum state
\beq
| \bc \rangle = D( \bc ) | 0 \rangle .
\label {defcohst}
\eeq 
By using the displacement relation (\ref {DaD}),
we may show that the coherent state is an \eis 
of every \ani \op $ a_n $:
\bea
a_n | \bc \rangle & = & a_n D( \bc ) | 0 \rangle 
= D( \bc ) \, D^\dgr( \bc ) a_n D( \bc ) | 0 \rangle \nn\\
& = & D( \bc ) \, \left( a_n + \c_n \right) | 0 \rangle
=  D( \bc ) \, \c_n | 0 \rangle
= \c_n \, D( \bc ) | 0 \rangle \nn\\
& = & \c_n | \bc \rangle .
\label {cseig}
\eea
By using the product formula (\ref {Dp})
for the displacement \opp,
we may write the coherent state in the form
\bea
| \bc \rangle & = & D( \bc ) | 0 \rangle 
= \prod_i \left[ 1 + \ad_i \c_i - \sc_i a_i
+ \left( \ad_i a_i - \thalf \right) \sc_i \c_i \right] | 0 \rangle \nn\\
& = & \prod_i \left( 1 + \ad_i \c_i - \thalf \sc_i \c_i \right) | 0 \rangle
\nn\\
& = & 
\exp \left( \sum_i \left( \ad_i \c_i - \thalf \sc_i \c_i \right) \right)
| 0 \rangle .
\label {pcohst}
\eea
It may be worth emphasizing that in this formula
the \cre \op $ \ad_i $ stands to the \emph{left} of the 
Grassmann number $ \c_i $.  Apart from these ordering
considerations, this formula takes a form closely analogous
to the one that defines bosonic coherent states.
\par
The adjoint of the coherent state $ | \bc \rangle $ is
\beq
\langle \bc | = \langle 0 | D^\dgr( \bc ) = \langle 0 | 
\exp \left( \sum_i \left( \cs_i a_i - \thalf \sc_i \c_i \right) \right) ,
\label {csad}
\eeq
and it obeys the relation
\beq
\langle \bc | \ad_n = \langle \bc | \cs_n .
\label {adstevrel}
\eeq
The inner product of two coherent states is
\beq
\langle \bc | \bb \rangle = 
\exp \left( \sum_i \left( \cs_i \b_i 
- \thalf ( \cs_i \c_i + \sb_i \b_i ) \right) \right) ,
\label {ip}
\eeq 
so that
\bea
\langle \bb | \bc \rangle
\, \langle \bc | \bb \rangle 
& = & \exp \left[ - \sum_i ( \sb_i - \cs_i ) ( \b_i - \c_i ) \right] \nn\\
& = & \prod_i \left[ 1 - ( \sb_i - \cs_i ) ( \b_i - \c_i ) \right] .
\label {sip}
\eea
\par
In contrast to the case of bosons,
we may for fermions define
for any set $ \ba = \{ \a_i \} $ of Grassmann numbers
the normalized \eis $ | \ba \rangle^\prime $
of the fermion creation operators $ \ad_i $
as the displaced state
\beq
| \ba \rangle^\prime = D( \ba ) | \bone \rangle
\label {fcohst}
\eeq
where $ | \bone \rangle $
is the state in which every mode is filled:
\beq
| \bone \rangle = \prod_n \ad_n | 0 \rangle .
\label {1}
\eeq
By using the displacement relation (\ref {DadD}),
we may  show that the state $ | \ba \rangle^\prime $
is an \eis of every \cre \op $ \ad_n $:
\bea
\ad_n | \ba \rangle^\prime & = & \ad_n D( \ba ) | \bone \rangle
= D( \ba ) \, D^\dgr( \ba ) \ad_n D( \ba ) | \bone \rangle \nn\\
& = & D( \ba ) \, \left( \ad_n + \sa_n \right) | \bone \rangle
=  D( \ba ) \, \sa_n | \bone \rangle
= \sa_n \, D( \ba ) | \bone \rangle \nn\\
& = & \sa_n | \ba \rangle^\prime .
\label {fseig}
\eea
The adjoint relation is
\beq
\, ^\backprime\langle \ba | a_n = \, ^\backprime\langle \ba | \a_n .
\label {adfseig}
\eeq
An explicit formula for the \eis $ | \ba \rangle^\prime $
follows from its definition (\ref {fcohst}):
\beq
| \ba \rangle^\prime = \prod_i 
\left( 1 - \sa_i a_i + \thalf \sa_i \a_i \right)
| \bone \rangle .
\label {expfcohst}
\eeq
 
\subsection{Intrinsic Descriptions of Fermionic States}
The occupation-number description
of states of fermions has well-known ambiguities.
For $ n \ne m $, for example,
the state $ | 1_n 1_m \rangle $ may be interpreted
as $ \ad_n \ad_m | 0 \rangle $
or as $ \ad_m \ad_n | 0 \rangle = - \ad_n \ad_m | 0 \rangle $.
\par
The creation operators themselves provide
an unambiguous description of fermionic states,
\beq
| \psi \rangle = 
\sum_{ \{n\} } c( n_1 , n_2 , \dots ) 
\, \ad_{n_1} \ad_{n_2} \dots \ad_{n_m} 
\, | 0 \rangle ,
\label {fstate}
\eeq
which transfers to the coherent-state representation
\beq
\langle \ba | \psi \rangle =
\exp \left( - \thalf \sum_n \sa_n \a_n \right)
\, \sum_{\{n\}} c( n_1 , n_2 , \dots ) \, \sa_{n_1} \sa_{n_2} \dots \sa_{n_m} 
\label {cfstate}
\eeq 
without any ambiguity or extra minus signs.
Because coherent states are defined in terms
of bilinear forms in anti-commuting variables,
there is no need to adopt a standard ordering of the modes.
\section{Grassmann Calculus}
\subsection{Differentiation}
Since the square of any Grassmann variable vanishes,
the most general function $ f( \x ) $
of a single anti-commuting variable
$ \x $ is linear in $ \x $
\beq
f( \x ) = u + \x t .
\label {fx}
\eeq
We define the left derivative of
the function $ f( \x ) $ with respect to
the Grassmann variable $ \x $ as
\beq
\frac { d f( \x ) }{ d\x } = t .
\label {df}
\eeq
Note that if the variable $ t $ is anti-commuting,
then we may also write the function $ f( \x ) $ in the form
\beq
f( \x ) = u - t \x .
\label {xf}
\eeq
Now to form the left derivative,
we first move $ \x $ past $ t $,
picking up a minus sign and obtaining
the form (\ref {fx}) and the result (\ref {df}). 
In this case, the right derivative is $ - t $.
In the present work, we shall use left derivatives
exclusively and shall refer to them simply
as derivatives.
\subsection{Even and Odd Functions}
It is useful to distinguish between
functions that commute
with Grassmann variables and ones that do not.
We shall say that a function $ f( \ba ) $
that commutes with Grassmann variables
is \emph{even\/}
and that a function $ f( \ba ) $
that anti-commutes with Grassmann variables
is \emph{odd\/}.
We shall often note the evenness or oddness
of the functions we introduce.

\subsection{Product Rule}
To compute the derivative of the product
of two functions $ f( \ba ) $ and $ g( \ba ) $
with respect to a particular variable $ \a_i $,
one may explicitly move the $ \a_i $ in
$ g( \ba ) $ through $ f( \ba ) $ or one may move
the operator representing differentiation
through the function $ f( \ba ) $.
In either case if the function $ f( \ba ) $
is odd, then one picks up a minus sign.
The product rule is thus
\beq 
\frac { \p }{ \p \a_i } \left( f( \ba ) \, g( \ba ) \right) =
\frac { \p f( \ba ) }{ \p \a_i } \, g( \ba )
+ \s( f ) \, f( \ba ) \, \frac { \p g( \ba )  }{ \p \a_i } ,
\label {prodrule}
\eeq
where the sign $ \s( f ) $ of $ f( \ba ) $
is $ - 1 $ if $ f( \ba ) $
is an odd function and $ + 1 $ if $ f( \ba ) $
is even.

\subsection{Integration}
We define a sort of integration over the complex
Grassmann variables by the following rules
\bea
\int \! d\a_n & = & \int \! d\sa_n = 0 \label {I0} \\
\int \! d\a_n \, \a_m & = & \d_{nm} \label {I1} \\
\int \! d\sa_n \, \sa_m & = & \d_{nm} .
\label {intdef}
\eea
This integration due to Berezin~\cite{b}
is exactly equivalent to left differentiation.
\par
We shall typically be concerned with
pairs of anti-commuting variables
$ \a_i $ and $ \sa_i $, and for such pairs
we shall adhere to the notation
\beq
\int \! d^2\a_n = \int d\sa_n \, d\a_n 
\label {d2a}
\eeq
in which the differential of the
conjugated variable comes first.
Note that 
\beq
d\a_n \, d\sa_n = - d\sa_n \, d\a_n .
\label {warnac}
\eeq
We have been using boldface type to denote 
sets of Grassmann variables;
we shall extend that use to write
multiple integrals over such sets in the succinct form
\beq
\int \! d^2\ba \equiv \int \prod_i d^2\a_i .
\label {dba}
\eeq
We shall also occasionally employ the concise notation
\beq
\bsa \cdot \bb \equiv \sum_n \sa_n \b_n 
\label {sumdot}
\eeq 
for sums of simple products over all the modes of the system.
\par
The simple integral formula 
\bea
\int \! d^2\a_n \, e^{ - \sa_p \a_q } 
& = & \int \! d\sa_n \, d\a_n \left( 1 - \sa_p \a_q \right)
= - \int \! d\sa_n \, d\a_n  \, \sa_p \a_q \nn\\
& = & \int \! d\sa_n \, \sa_p \, d\a_n \, \a_q
= \int \! d\sa_n \, \sa_p \, \d_{ n q } = \d_{ n p } \, \d_{ n q }  
\label {ex0}
\eea
provides a useful example of Grassmann integration.
We also note the general rule
\beq
\int \! d^2\a \, f( \lambda \a ) =
| \lambda |^2 \int \! d^2 \b \, f( \b ) ,
\label {ex1}
\eeq
in which $ \lambda $ is an arbitrary complex number
and in which $ f ( \lambda \a ) $ is an abbreviation
for a function which necessarily depends on both $ \lambda \a $
and $ \lambda^* \a^* $. 
This rule owes its strange appearance to the definition
of integration as differentiation.
\par
Some further examples are the integral of 
the exponential function 
\beq
\int \! d^2\a \, \exp \left( \sb \a + \sa \c + \a \sa \right)
= \exp \left( \sb \c \right) 
\label {ex2}
\eeq
and the Fourier transform of a gaussian
\beq
\int \! d^2\x \, \exp \left( \a \xs - \x \sa + \l \x \xs \right)
= \l \, \exp \left( \frac{ \a \sa }{ \l } \right)
\label {ex3}
\eeq
where $ \l $ is an arbitrary complex number.
The latter integral can be written in a somewhat 
more-general form which is no longer a Fourier transform:
\beq
\int \! d^2\x \, \exp \left( \a \xs - \x \sb + \l \x \xs \right)
= \l \, \exp \left( \frac{ \a \sb }{ \l } \right) .
\label {ex4}
\eeq

\subsection{Integration by Parts}
Let us first observe that the 
integral of a derivative vanishes
\beq
\int \! d^2\ba \, \frac { \p f( \ba ) }{ \p \a_i } = 0 
\label {intofder0}
\eeq
because the derivative  with respect to 
the variable $ \a_i $ lacks the variable $ \a_i $.
In particular the integral of the derivative 
of the product of two functions also vanishes,
and so by using the product rule (\ref {prodrule}),
we have
\beq
\int \! d^2\ba \, \frac { \p }{ \p \a_i } \left( f( \ba ) 
\, g( \ba ) \right) = \int \! d^2\ba \, \left[
\left( \frac { \p f( \ba ) }{ \p \a_i } \right)
\, g( \ba ) + \s( f ) \,  f( \ba ) \frac { \p g( \ba ) }{ \p \a_i } 
\right] = 0 ,
\label {intofderp}
\eeq
which is the formula for integration by parts,
\beq
\int \! d^2\ba \, \left( \frac { \p f( \ba ) }{ \p \a_i } \right)
\, g( \ba ) =
- \s( f ) \, \int \! d^2\ba \,  f( \ba ) \,
\frac { \p g( \ba ) }{ \p \a_i } ,
\label {intbyparts}
\eeq
where the sign $ \s( f ) $ is $ +1 $ 
if the function $ f( \ba ) $ is even
and $ -1 $ if it is odd.

\subsection{Completeness of the Coherent States}
We may use our \G calculus to show that the
\costs are complete.  Let us consider the state 
\beq
| f \rangle = \left( c + d \ad \right) | 0 \rangle ,
\label {arbst}
\eeq
which for arbitrary complex numbers $ c $ and $ d $
is an arbitrary single-mode state.
Then its inner product $ \langle \c | f \rangle $
with the \cost $ | \c \rangle $ is the correct
weight function for the \co_st expansion
since
\beq
\int \! d^2\c \langle \c | f \rangle | \c \rangle =
\int \! d^2\c ( c + d \cs ) 
\, \left( 1 + \c \cs - \c \ad \right) | 0 \rangle 
= \left( c + d \ad \right) | 0 \rangle
= | f \rangle .
\label {co_stcompl}
\eeq
The reader may generalize this example to the
multi-mode case.  The coherent states in fact
are over-complete.

\subsection{Completeness of the Displacement Operators}
For a single mode, the identity operator $ I $ 
and the traceless \ops $ a, \ad $, 
and $ \thalf - \ad a $ form a complete set of \opsp.
Since by using the expression (\ref {Dp})
and our \G calculus,
we may write each of these \ops
as an integral over the \disp \ops
\bea
I & = & \int \! d^2\c \, \c \cs \, D( \c ) \label {DI} \\
a & = & \int \! d^2\c \, ( - \c ) \, D( \c ) \label {Da} \\
\ad & = & \int \! d^2\c \, \cs \, D( \c ) \label {Dad} \\
\thalf - \ad a & = & \int \! d^2\c \, D( \c ) ,
\label {Dada}
\eea
it follows that the \disp \ops
form a complete set of \ops for that mode.
It is easy to generalize this proof to the
multi-mode case.  The \disp \ops are over-complete.

\section{Operators}
Some operators can be written as sums of products 
of even numbers of \cre and \ani \opsp;
we shall call such
operators \emph{even}.
Operators that can be written as sums of products 
of odd numbers of \cre and \ani \opsp
we shall shall speak of as \emph{odd}.
Although most operators are neither even nor odd,
the operators of physical interest are either
even or odd.
The number \op $ \ad a $, for example,
is even, while the \cre and \ani \opsp, $ \ad $ and $ a $,
are odd. 
\par
The operators of quantum mechanics
and of quantum field theory
do not themselves involve Grassmann variables.
Thus even operators commute with
Grassmann variables, 
while odd ones anti-commute.

\subsection{The Identity Operator}
If we compare the integral
\beq
\int \! d^2\ba \, | \ba \rangle \langle \ba | \bb \rangle 
= \int \! d^2\ba \, \exp \left( \sum_i \left(
\ad_i \a_i + \sa_i \b_i + \a_i \sa_i 
+ \thalf \b_i \sb_i \right) \right) | 0 \rangle
\label {id1}
\eeq 
with the integral-formula (\ref {ex2})
and identify $ \sb $ and $ \c $ in that formula
with $ \ad $ and $ \b $ in this integral,
then we have
\beq
\int \! d^2\ba \, | \ba \rangle \langle \ba | \bb \rangle
= \exp \left( \sum_i \left(
\ad_i \b_i + \thalf \b_i \sb_i \right) \right) | 0 \rangle
= | \bb \rangle .
\label {id2}
\eeq
Since the coherent states form a complete set
of states, as shown by the expansion (\ref {co_stcompl}),
it follows that
the identity operator is given by the integral
\beq
I = \int \! d^2 \ba \, | \ba \rangle \langle \ba | .
\label {I}
\eeq 
\par
The corresponding expression for the identity operator
in terms of the \eiss $ | \ba \rangle^\prime $
of the \cre \ops is 
\beq
I = \int \! \prod_i \left( - d^2 \a_i \right) \, 
| \ba \rangle^\prime \, ^\backprime\langle \ba | .
\label {I'}
\eeq

\subsection{The Trace}
The trace of an arbitrary \op $ B $
is the sum of the diagonal matrix
elements of $ B $ in the $n$-quantum states,
\beq
\Tr B = \sum_n \langle n | B | n \rangle ,
\label {sumtr}
\eeq
which shows that the trace of an operator 
that is odd vanishes.
By inserting the preceding formula (\ref {I})
for the identity \opp, we have
\beq
\Tr B = \sum_n \int \! d^2 \a \, \langle n | \a \rangle
\langle \a | B | n \rangle .
\label {sumtr2}
\eeq
If we move the coherent-state matrix element
$ \langle n | \a \rangle $ to the right
of the matrix element of the even \op $ B $,
then we see from the formula (\ref{pcohst})
that minus signs arise that can be absorbed
into the argument of either of the two coherent states,
\beq
\Tr B = \sum_n \int \! d^2 \a \, 
\langle \a | B | n \rangle \langle n | - \a \rangle 
= \sum_n \int \! d^2 \a \, 
\langle - \a | B | n \rangle \langle n | \a \rangle ,
\label {sumtr3}
\eeq
in which the sum $ \sum_n | n \rangle \langle n | = I $
is the identity \opp.
The resulting multi-mode trace formula is
\beq
\Tr B = \int \! d^2 \ba 
\, \langle \ba | B | - \ba \rangle 
= \int \! d^2 \ba \, \langle - \ba | B | \ba \rangle ,
\label {trace}
\eeq
which holds also for odd operators, both sides vanishing.
An important example is the trace of the dyadic
operator $ | \bb \rangle \langle \bc | $,
\bea
\Tr | \bb \rangle \langle \bc | & = &
\int \! d^2 \ba \, \langle \ba | \bb \rangle \langle \bc | - \ba \rangle
= \int \! d^2 \ba \, \langle \bc | - \ba \rangle \langle \ba | \bb \rangle 
\nn\\
& = & \int \! d^2 \ba \, \langle - \bc | \ba \rangle \langle \ba | \bb \rangle 
= \langle - \bc | \bb \rangle = 
\langle \bc | - \bb \rangle ,
\label {trex}
\eea
in which we have used the completeness relation (\ref {I}).
Since the coherent states are complete,
we may replace in this formula either the ket $ | \bb \rangle $
or the bra $ \langle \bc | $ 
with its image $ F | \bb \rangle $ 
or $ \langle \bc | F $ under the action of the
arbitrary \op $ F $ and obtain the trace formula
\beq
\Tr \left( F | \bb \rangle \langle \bc | \right) = 
\Tr \left( | \bb \rangle \langle \bc | F \right) =
\langle - \bc | F | \bb \rangle = \langle \bc | F | - \bb \rangle .
\label {trFdy}
\eeq
 
\subsection{Physical States and Operators}
A state $ | \psi \rangle $ is \emph{physical\/}
if it changes at most by a phase
when subjected to a rotation of angle $ 2 \pi $
about any axis,
\beq
U( \hat n , 2 \pi ) | \psi \rangle = e^{ i \theta } 
\, | \psi \rangle .
\label {physst}
\eeq
Since fermions carry half-odd-integer spin,
a state of one fermion or of any odd number
of fermions changes by the phase factor $ -1 $.  
States that contain no fermions or only even
numbers of fermions are invariant under such
$ 2 \pi $ rotations.
\par
Thus physical states are linear combinations
of states with odd numbers of fermions
or linear combinations of states with even
numbers of fermions.  But a state that is
a linear combination of a state that contains
an odd number of fermions and another that 
contains an even number of fermions is excluded.  
For instance, the state 
\beq
\frac { 1 } { \sqrt{2} }
\, \left( | 0 \rangle + | 1 \rangle \right)
\label {sick}
\eeq
is unphysical because under a $ 2 \pi $ rotation
it changes into a different state:
\beq
U( \hat n , 2 \pi ) \frac { 1 } { \sqrt{2} }
\, \left( | 0 \rangle + | 1 \rangle \right)
= \frac { 1 } { \sqrt{2} } \,
\left( | 0 \rangle - | 1 \rangle \right)
\ne e^{ i \theta } \frac { 1 } { \sqrt{2} }
\, \left( | 0 \rangle + | 1 \rangle \right) .
\label {disease}
\eeq
\par
We define an operator as \emph{physical\/}
if it maps physical states onto physical states.
Physical operators are either even or odd. 
\par
In all physical contexts that have been explored experimentally, 
the number of fermions (or more generally the number 
of fermions minus the number of anti-fermions) is strictly conserved.
That conservation law leads to certain further 
restrictions on the permissible states of the field.
If we let $ N = \sum_k \ad_k a_k $ be the fermion number,
the law requires that any state arising from 
an eigenstate of $ N $ must remain an eigenstate of $ N $.
This law can be derived from an assumed $ U(1) $ invariance
of all the interactions under the transformation
$ U( \th ) = \exp \left( i \th N \right) $,
which changes $ a $ and $ \ad $ to
\beq
e ^ { - i \th N } a e ^ { i \th N } = e ^ { i \th } a 
\label {aq}
\eeq
and
\beq
e ^ { - i \th N } \ad e ^ { i \th N } = e ^ { - i \th } \ad .
\label {adq}
\eeq
Fermion conserving interactions involving the $ a_k $ 
and $ \ad_k $ are ones in which the phase factors
$ e ^ { \pm i \th } $ all cancel.
If a system begins in a state with a fixed number of fermions,
the conservation law restricts the set of accessible states
considerably more than the $ 2 \pi $ super-selection rule 
mentioned earlier.  
Transitions can not be made, for example,
between states with 
different even fermion numbers or between states with 
different odd fermion numbers.

\subsection{Physical Density Operators} 
A physical density operator can be written
as a sum of dyadics of physical states 
with positive coefficients that add up to unity.
It follows that a physical density operator $ \rho $
is a positive hermitian operator of unit trace:
for any state $ | \psi \rangle $ 
\bea
\langle \psi | \rho | \psi \rangle & \ge & 0 \\
\rho ^\dgr & = & \rho \\
\Tr \rho & = & 1 .
\label {rho}
\eea
\par
Physical density operators are invariant under a $ 2 \pi $ rotation.
Thus the one-mode operator 
\beq
\rho = \half \left( | 0 \rangle \langle 0 |
+ | 1 \rangle \langle 1 | \right) ,
\label {pdo1}
\eeq
for example,
is a physical density operator,
but the dyadic
\beq
\half \, \left( | 0 \rangle + | 1 \rangle \right)
\, \left(  \langle 0 | + \langle 1 | \right)
\label {sdo1}
\eeq
is not. 
In this work we shall consider only
density operators that are physical in this sense.
\par
The dynamical problems we solve do not always
begin with a fixed number of fermions.
More generally they begin with a mixture of states
with different fermion numbers, that is
with density operators of the form
\beq
\rho = \sum_ { N' } p_ { N' } \, | N' \rangle \langle N' | 
\label {rhoN'}
\eeq
where the $ p_ { N' } $ are real and non-negative.
Such density operators are invariant under
the transformation $ U( \th ) = \exp ( i \th N ) $,
and the fermion conservation law assures us that 
they will always remain so,
\beq
e ^ { - i \th N } \rho e ^ { i \th N } = \rho .
\label {UrUd}
\eeq
\par
The coherent states do undergo a simple
change under this transformation,
\beq
U( \th ) | \a \rangle = e ^ { i \th N } | \a \rangle 
= | e ^ { i \th } \a \rangle 
\label {ath}
\eeq
\beq
\langle \a | U^\dgr( \th ) = \langle \a | e ^ { - i \th N } 
= \langle e ^ { i \th } \a | ,
\label {adth}
\eeq
which leaves their scalar product invariant,
\beq
\langle e ^ { i \th } \a | e ^ { i \th } \a \rangle 
= \langle \a | \a \rangle .
\label {spinv}
\eeq

\section{Delta Functions and Fourier Transforms}

We can define a function
\bea
\d ( \bx - \bz ) & \equiv & 
\int \! d^2 \ba \,
\exp \left( \sum_n \left(
\a_n \left( \xs_n - \sz_n \right)
- \left( \x _n - \z_n \right) \sa_n \right) \right) \label {Delta} \\
& = & \mbox{}
\prod_n 
\left( \x_n - \z _n \right)
\, \left( \xs _n - \sz _n \right) 
\label {delta}
\eea
which plays the role of a Dirac delta function
in that if $ f ( \bx ) $
is any function of the set $ \bx $ of 
Grassmann variables $ \{ \x_1, \x_2, \dots \} $,
then
\beq
\int \! d^2 \bx \, \d ( \bx - \bz ) \, f ( \bx ) 
= f ( \bz ) . 
\label {why delta}
\eeq 
The delta function is doubly even: it commutes with
\G numbers and $ \d ( \x - \z ) = \d ( \z - \x ) $.
\par
We have been using the term Fourier transform
to denote an integral of the form
\beq
\tilde f ( \a ) = \int \! d^2\x \, e^{ \a \xs - \x \sa } f ( \x ) .
\label {defFT}
\eeq
The delta-function identity (\ref {Delta})
implies that the inverse Fourier transform
is given by the similar formula
\beq
f ( \x ) = \int \! d^2\a \, e^{ \x \sa - \a \xs } \tilde f ( \a ) .
\label {defIFT}
\eeq
The identity (\ref {Delta}) also leads to two forms
of Parseval's relation:
\beq
\int \! d^2\a \, \tilde f ( \a ) \, \left[ \tilde g( \a ) \right]^*
= \int \! d^2\x \, f ( \x ) \, g^*( \x ) 
\label {Par1}
\eeq
and
\beq
\int \! d^2\a \, \tilde f ( \a ) \, \tilde g( - \a ) 
= \int \! d^2\x \, f ( \x ) \, g( \x ) ,
\label {Par2}
\eeq
which apply also to operator-valued functions 
provided that complex conjugation is replaced by
hermitian conjugation.
\par
We may use the formula (\ref {Delta}) for the delta function
to derive a fermionic analog of the convolution theorem:
\bea
\int \! d^2 \x \, 
e^{ \a \xs - \x \sa } 
f( \x ) g ( \x ) 
\! & = \! & \int \! d^2\b \,  d^2 \x \,
e^{ ( \a - \b ) \xs - \x ( \sa - \sb ) }
f( \x ) \, \int \! d^2 \eta e^{ \b \eta^* - \eta \sb }g ( \eta ) \nn\\
\! & = & \! \int \! d^2\b \, \tilde f( \a - \b ) \, \tilde g( \b ) ,
\label {convo}
\eea
which expresses the Fourier transform of the product
of the two functions $ f( \x ) $ and $ g ( \x ) $ as
the convolution of their Fourier transforms
$ \tilde f( \a - \b ) $ and $ \tilde g( \b )$. 
\par
By using 
the normally ordered form (\ref {DN}) of the displacement operator, 
the eigenvalue property of the coherent states,
and the preceding formula (\ref {Delta}) for the delta function
we find
\bea
\int \! d^2\c \, \langle \c | D( \a ) | \c \rangle
& = & \int \! d^2\c \, 
\langle \c | e^{ \ad \a }
\, e^{  - \sa a } | \c \rangle
\, e ^{ \shalf \a \sa } \nn\\
& = & \int \! d^2\c \,
e^{ \sc \a - \sa \c +  \shalf \a \sa } \nn\\
& = & \d( \a ) e ^{ \shalf \a \sa } = \d( \a ) .
\label {del1} 
\eea
The addition rule (\ref {DD}) for successive displacements
now implies that for the multi-mode case
\beq
\int \! d^2\bc \, \langle \bc | D( \ba ) D( - \bb ) | \bc \rangle
= \d ( \ba - \bb ) .
\label {del2}
\eeq

\section{Operator Expansions}
\par
The preceding delta-function identity (\ref {del2})
and the completeness (\ref {DI}--\ref {Dada})
of the displacement \ops give us a means of expanding 
an arbitrary operator $ F $ in the form
\beq
F = \int \! d^2\bx f ( \bx ) D( -\bx ) .
\label {exp1}
\eeq
We may solve for the weight function $ f ( \bx ) $
by multiplying on the right
by the displacement \op $ D( \ba ) $
and then taking the diagonal coherent-state matrix element
in the state $ | \bb \rangle $ and integrating over $ \bb $:
\bea
\int \! d^2\bb \, \langle \bb | F D( \ba ) | \bb \rangle & = &
\int \! d^2\bb \, \int \! d^2\bx f ( \bx ) \,
\langle \bb | D( -\bx ) D( \ba ) | \bb \rangle \nn\\
& = & \int \! d^2\bx f ( \bx ) \, \d( \ba - \bx ) 
= f ( \ba ) .
\label {delta_identity}
\eea
The full expansion is thus 
\beq
F = \int \! d^2\bx \, \int \! d^2\bb \, 
\langle \bb | F D( \bx ) | \bb \rangle \,
D( -\bx ) .
\label {delta_expansion}
\eeq
Such expansions will prove useful in the sections that follow.
\par
The formula (\ref {Delta}) for the delta
function $ \d ( \x - \z ) $ may be interpreted
as a trace identity.
From the eigenvalue property of the coherent states,
it follows that
\beq
\d ( \x - \z ) = \int \! d^2 \a \,
e ^ { \a \xs - \x \sa } \, e ^ { \z \sa - \a \zs }
= \int \! d^2 \a \,
e ^ { \a \xs - \x \sa } \, \langle \a | 
e ^ { \z \ad } e ^ { - a \zs } | \a \rangle
\label {deqee}
\eeq 
in which we recognize
the normally ordered form (\ref {DN}) of the displacement operator 
\beq
\d ( \x - \z ) = \int \! d^2 \a \,
e ^ { \a \xs - \x \sa } \, \langle \a | D_N( \z ) | \a \rangle .
\label {deqDN}
\eeq
By using the trace formula (\ref {trFdy}),
we may write this delta function as the trace 
\bea
\d ( \x - \z ) & = & \int \! d^2 \a \,
e ^ { \a \xs - \x \sa } \, 
\Tr \left[ D_N( \z ) | \a \rangle \langle - \a | \right] 
\nn\\
& = & \Tr \left[ D_N( \z ) E_A( - \x ) \right]
\label {deqtrDNEA}
\eea
The subscript $ A $ has been chosen to indicate its anticipated
of the product of the normally ordered displacement operator $ D_N( \z ) $
with an even operator $ E_A( \x ) $
defined as the Fourier transform
\beq
E_A( \x ) = \int \! d^2 \a \,
e ^ { \x \sa - \a \xs } \,
| \a \rangle \langle - \a | 
\label {EA}
\eeq
of the \co_st dyadic $ | \a \rangle \langle - \a |$.
As intimated by its subscript,
the operator $ E_A( \x ) $ will turn out 
to be useful for dealing with anti-normally ordered operators.
\par
We may now use the completeness (\ref {DI}--\ref {Dada})
of the \disp \ops and the trace identity (\ref {deqtrDNEA}) 
to expand an arbitrary \op $ F $ in terms of the
normally ordered \disp \ops $ D_N( \x ) $,
\beq
F = \int \! d^2\x \, f ( \x ) D_N( - \x ) .
\label {FfDN}
\eeq
We may solve for the function $ f ( \x ) $ by multiplying
on the right
by the \op $ E_A( \z ) $ and forming the trace:
\beq
\Tr \left[ F E_A( \z ) \right] =
\int \! d^2\x \, f ( \x ) \Tr \left[ D_N( - \x ) E_A( \z ) \right] 
= \int \! d^2\x \, f ( \x ) \d( \z - \x ) = f ( \z ) .
\label {f(z)}
\eeq
The full expansion is thus
\beq
F = \int \! d^2\bx \, \Tr \left[ F E_A( \bx ) \right]
\, D_N( - \bx ) .
\label {FTrDN}
\eeq 
\par
By using the \G calculus,
one may compute the Fourier transform
(\ref {EA}) of the \co_st dyadic $ | \a \rangle \langle - \a |$
and find for the \op $ E_A( \x ) $ the formulas
\bea
E_A( \x ) & = & | 0 \rangle \langle 0 |
- ( \xs + \ad ) | 0 \rangle \langle 0 | ( \x + a ) \\
& = & 2 ( \thalf - \ad a )
+ \x \xs a \ad + \x \ad - \xs a 
\label {EAada}
\eea
with which it is easy to exhibit 
the completeness of the \ops $ E_A( \x ) $:
\bea
I & = & \int \! d^2\x \,\, 2 ( 1 + \xs \x ) \, E_A( \x ) \\
a & = & \int \! d^2\x \,\, ( - \x ) \, E_A( \x ) \\
\ad & = & \int \! d^2\x \,\, ( - \xs ) \, E_A( \x ) \\
\thalf - \ad a & = & \int \! d^2\x \,\, \thalf \x \xs \, E_A( \x ) .
\label {EAcompl}
\eea
\par
Since the \ops $ E_A( \x ) $ are complete,
we may expand an arbitrary \op $ G $ in terms of them,
\beq
G = \int \! d^2\x \, g( \x ) E_A( - \x ) 
\label {GEA}
\eeq
and then use the trace formula (\ref {deqtrDNEA})
and the evenness of the displacement operators
to evaluate the weight function $ g( \x ) $,
\beq
\Tr \left[ D_N( \z ) G \right] = 
\int \! d^2\x \, g( \x ) \, \Tr \left[ D_N( \z ) E_A( - \x ) \right]
= \int \! d^2\x \, g( \x ) \, \d( \x - \z ) 
= g( \z ) .
\label {g}
\eeq
The full expansion is thus
\beq
G = \int \! d^2\bx \, \Tr \left[ G D_N( \bx ) \right] \,
E_A( - \bx ) .
\label {EAexp}
\eeq

\section{Characteristic Functions}
For a system described by the density operator $ \rho $,
we define the characteristic function $ \chi ( \bx ) $
of Grassmann argument $ \bx $ ( and $ \bxs $ )
as the mean value 
\beq
\chi ( \bx ) = \Tr \left[ 
\rho \, \exp \left( \sum_n \left( \x_n a^\dgr_n - a_n \xs_n \right) \right)
\right] .
\label {char}
\eeq 
It is thus a species of Fourier transform
of the density \op $ \rho $.
Because $ \x_i^2 = \xi_i^{*2} = 0 $,
we may expand the exponential as
\beq
\chi ( \bx ) = \Tr \left[
\rho \, \prod_n \left(
1 + \x_n a^\dgr_n - a_n \xs_n 
+ \xs_n \x_n ( a^\dgr_n a_n - \thalf )
\right)
\right] .
\label {charp}
\eeq
\par
We may also define the normally ordered
characteristic function $ \chi_N ( \bx ) $ as
\beq
\chi_N ( \bx ) = \Tr \left[
\rho \, \exp\left( \sum_n \x_n a^\dgr_n \right)
\, \exp \left( - \sum_m a_m \xs_m \right)
\right] 
\label {charn}
\eeq
with the expansion
\beq
\chi_N ( \bx ) = \Tr \left[
\rho \, \prod_n \left(
1 + \x_n a^\dgr_n - a_n \xs_n
+ \xs_n \x_n a^\dgr_n a_n 
\right)
\right] .
\label {charnp}
\eeq
\par
The anti-normally ordered
characteristic function $ \chi_A ( \bx ) $ is 
\bea
\chi_A ( \bx ) & = & \Tr \left[
\rho 
\, \exp\left( - \sum_m a_m \xs_m \right)
\, \exp\left( \sum_n \x_n a^\dgr_n \right)
\right] \label {charA}\\
& = &
\Tr \left[
\rho \, \prod_n \left(
1 + \x_n a^\dgr_n - a_n \xs_n
+ \xs_n \x_n ( a^\dgr_n a_n - 1 )
\right)
\right] .
\label {chara}
\eea
Because the density \op $ \rho $ 
is an even \op and because
the displacement \ops are constructed
from bilinear forms in fermionic quantities, 
it follows that the characteristic functions
are even in the sense that they commute
with \G variables.

\subsection{The S-Ordered Characteristic Function}
We may define a more general ordering
of the \ani \op $ a_n $ and the \cre \op $ a^\dgr_n $,
much as we did earlier for boson-field operators~\cite{kcrg1}.
It is an ordering specified by a real parameter $ s $
that runs from $ s = - 1 $ for anti-normal ordering
to $ s = 1 $ for normal ordering.
For the quadratic case, the s-ordered product 
for fermions is
\beq
\{ a^\dgr_n a_n \}_s = a^\dgr_n a_n + \thalf ( s - 1 ) ,
\label {adas}
\eeq
to which we append the trivial definitions
\beq
\{ a^\dgr_n \}_s = a^\dgr_n 
\and
\{ a_n \}_s = a_n .
\label {trivdef}
\eeq
We note that the definition (\ref {adas})
differs by a crucial sign from that~\cite{kcrg1}
of $s$-ordering for bosonic \ops
$ b_n $ and $ b^\dgr_n $:
\beq
\{ b^\dgr_n b_n \}_s = b^\dgr_n b_n + \thalf ( 1 - s ) .
\label {adasb}
\eeq
\par
In particular the anti-normally ordered product
$ \{ a^\dgr_n a_n \}_{-1} $ is $ \mbox{} - a_n a^\dgr_n $,
and the symmetrically ordered product
$ \{ a^\dgr_n a_n \}_0 $ is half the commutator,
\beq
 \{ a^\dgr_n a_n \}_0 = \thalf \left[ a_n , \ad_n \right] .
\label {aa0}
\eeq
We define the s-ordered characteristic function
$ \chi ( \bx , s ) $ as
\bea
\chi ( \bx , s ) & = & 
\Tr \left[ \rho
\, \left\{
\, \exp \left( \sum_n \left( \x_n a^\dgr_n - a_n \xs_n \right) \right)
\right\}_s
\right] \\
& = &
\Tr \left[
\rho \, \prod_n \left(
1 + \x_n a^\dgr_n - a_n \xs_n
+ \xs_n \x_n \, \{ a^\dgr_n a_n \}_s 
\right)
\right] \\
& = &
\Tr \left[
\rho \, \prod_n \left(
1 + \x_n a^\dgr_n - a_n \xs_n
+ \xs_n \x_n \, ( a^\dgr_n a_n + \thalf ( s - 1 ) )
\right)
\right] \\
& = & \Tr \left[
\rho
\, \exp \left( \sum_n \left( \x_n a^\dgr_n - a_n \xs_n 
+ \frac{ s }{ 2 } \, \xs_n \x_n \right) \right)
\right] \\
& = & \chi ( \bx ) \, 
\exp\left( \frac{ s }{ 2 } \, \sum_n \xs_n \x_n \right) ,
\label {chars}
\eea
which, incidentally, shows it to be an even function.
\par
A particularly useful example of these characteristic functions
is the case of the anti-normally ordered function
$ \chi_A ( \bx ) = \chi( \bx , -1 ) $.
We see by inserting the resolution (\ref {I})
of the identity between the exponential functions
in its definition (\ref {charA}) that 
\beq
\chi ( \bx , -1 ) = \Tr \left[
\rho
\, \exp\left( - \sum_m \b_m \xs_m \right)
\int \! d^2 \bb \, | \bb \rangle \langle \bb |
\, \exp\left( \sum_n \x_n \sb_n \right)
\right] ,
\eeq
in which we have replaced the \ani and \cre \ops
by their eigenvalues in the coherent states.
By using the trace formula (\ref {trex}), we find
\beq
\chi ( \bx , -1 ) = \int \! d^2 \bb \, 
\exp\left( \sum_n \left( \x_n \sb_n - \b_n \xs_n \right) \right) \,
\langle \bb | \rho | - \bb \rangle ,
\label {chiaft}
\eeq
which expresses the anti-normally ordered characteristic function
$ \chi ( \bx , -1 ) $ as the Fourier transform of the matrix element 
$ \langle \bb | \rho | - \bb \rangle $.
\par
If we define the s-ordered displacement operator
$ D ( \bx , s ) $ as  
\beq
D ( \bx , s ) = \left\{ D ( \bx ) \right\}_s 
= D ( \bx ) \exp \left( \frac{ s }{ 2 }
\sum_n \xs_n \x_n \right) ,
\label {Dbxs}
\eeq
then we may write the s-ordered characteristic function (\ref {chars}) 
as the trace
\beq
\chi ( \bx , s ) = \Tr \left[
\rho
\, D ( \bx , s ) \right] .
\label {chibxsDbxs}
\eeq

\section{S-Ordered Expansions for Operators}
A convenient extension of 
the definition of the operator $ E_A( \x ) $ is 
\beq
E( \bx , s ) \equiv E_A( \bx ) \, \exp \left( \frac { s + 1 }{ 2 }
\sum_n \xs_n \x_n \right) ,
\label {Exs}
\eeq
from which we note that 
\beq 
E_A( \bx ) = E( \bx , -1 ). 
\label {EAEx-1}
\eeq
This is one sense in which the operator $ E_A( \x ) $
is related to anti-normal ordering.
\par
By using the s-ordered operators $ D( \x , s ) $
and $ E( \x , s ) $,
we may generalize the expansions (\ref {FTrDN})
and (\ref {EAexp})
of the arbitrary operators $ F $ and $ G $ to
\bea
F & = & \int \! d^2\bx \, \Tr \left[ F E( \bx , -s ) \right]
\, D( - \bx , s ) \label {FTrDs} \\
G & = & \int \! d^2\bx \, \Tr \left[ G D( \bx , -s ) \right] 
\, E( - \bx , s ) \label {GTrEs} .
\eea
The obvious generalization
\beq
\d ( \bx - \bz ) = \Tr \left[ D( \bx , s ) E( - \bz, -s ) \right]
\label {DsEsdelta}
\eeq
of the trace formula (\ref {deqtrDNEA})
then gives the trace of the product $ FG $ as
\beq
\Tr \left[ F \, G \right] = 
\int \! d^2\bx \, \Tr \left[ F E( \bx , - s ) \right]
\, \Tr \left[ G D( - \bx , s ) \right] .
\label {TrFG}
\eeq
\par
We may now use the second Parseval relation (\ref {Par2})
to cast the expansions (\ref {FTrDs}) and (\ref {GTrEs})
into forms that will prove to be quite useful.
First let us define the complete sets of \ops 
$ \tilde D( \ba , s ) $ and $ \tilde E( \ba , s ) $ 
as the Fourier transforms of the \ops
$ D( \bx , s ) $ and $ E( \bx , s ) $:
\bea
\tilde D( \ba , s ) & \equiv & \int \! d^2\bx \, 
\exp \left( \sum_n
\left( \a_n \xs_n - \x_n \sa _n \right) \right) \,
D( \bx , s ) \label {FTD} \\
\tilde E( \ba , s ) & \equiv & \int \! d^2\bx \,
\exp \left( \sum_n
\left( \a_n \xs_n - \x_n \sa _n \right) \right) \,
E( \bx , s ) . \label {FTE}
\eea
Next let us define the weight functions
$ F_E ( \ba , - s ) $ and $ G_D ( \ba , -s ) $ 
as the Fourier transforms of the traces
\bea
F_E ( \ba , - s ) & \equiv & \int \! d^2\bx \,
\exp \left( \sum_n
\left( \a_n \xs_n - \x_n \sa _n \right) \right) \,
\Tr \left[ F E( \bx , - s ) \right] \label {F_E} \\
G_D ( \ba , -s ) & \equiv & \int \! d^2\bx \,
\exp \left( \sum_n
\left( \a_n \xs_n - \x_n \sa _n \right) \right) \,
\Tr \left[ G D( \bx , - s ) \right] .
\label {G_D}
\eea
It follows then from the Parseval relation (\ref {Par2})
and from the expansions (\ref {FTrDs}) and (\ref {GTrEs})
that the operators $ \tilde D( \ba , s ) $ and $ \tilde E( \ba , s ) $
form complete sets of \ops and afford us the 
expansions
\bea
F & = & \int \! d^2\ba \, F_E ( \ba , - s )
\, \tilde D( \ba , s ) \label {FTrFDs} \\
G & = & \int \! d^2\ba \, G_D ( \ba , -s ) 
\, \tilde E( \ba , s ) \label {GTrFEs} 
\eea
of the arbitrary operators $ F $ and $ G $\@.
Applying the Parseval relation (\ref {Par2})
to the trace formula (\ref {TrFG}), we have the 
trace relation
\beq
\Tr \left[ F \, G \right] =
\int \! d^2\ba \, F_E ( \ba , - s ) \, G_D ( \ba , s ) .
\label {TrFGa}
\eeq
\par
The \ops $ \tilde E( \ba , s ) $ are particularly simple
when $ s = \pm 1 $\@.
It follows from
the definitions (\ref {Exs}) and
(\ref {EA}) of the \ops $ \tilde E( \ba , s ) $
and $ E_A( \x ) $, and 
from the formula (\ref {Delta}) for the delta function
that the \op $ \tilde E( \ba , -1 ) $ is
just the \co_st dyadic
\beq
\tilde E( \ba , -1 ) = | \ba \rangle \langle - \ba | .
\label {E-1}
\eeq
Similarly, by using the definitions 
(\ref {Exs}) and (\ref {EA})
and the Fourier-transform relation (\ref {ex3}),
one may write the  \op $ \tilde E( \ba , 1 ) $ as the integral
\beq
\tilde E( \ba , -1 ) = \int \! \prod_i \left( - d^2 \b_i \right)
\, e^{ - ( \ba - \bb ) \cdot ( \bsa -\bbs ) } \,
| \bb \rangle \langle - \bb | .
\label {E(+1)1}
\eeq
By performing the integration and referring to
the explicit formula (\ref {expfcohst}),
we may show that the operator $ \tilde E( \ba , 1 ) $ is
the dyadic of the \eiss (\ref {fcohst}) of the \cre \ops 
$ | \ba \rangle^\prime  $:  
\beq
\tilde E( \ba , 1 ) = | \ba \rangle^\prime \, ^\backprime\langle - \ba | .
\label {E+1}
\eeq

\section{Quasi-Probability Distributions}
\par
Among the most important
of the foregoing expansions,
is the expansion (\ref {GTrEs})
when the \op $ G $ is the density \op $ \rho $,
\beq
\rho = \int \! d^2\bx \, \Tr \left[ \rho D( \bx , s ) \right]
\, E( - \bx , - s ) ,
\label {G->rho}
\eeq
in which case the trace is the s-ordered \char function
$ \chi( \bx , s ) $,
\beq
\rho = \int \! d^2\bx \, \chi( \bx , s )
\, E( - \bx , - s ) .
\label {rhochi}
\eeq
\par
We may define the s-ordered quasi-probability
distribution $ W( \ba , s ) $ as the Fourier
transform of the s-ordered characteristic function
$ \chi ( \bx , s ) $
\beq
W( \ba , s ) = \int \! d^2\bx \exp \left( \sum_n
\left( \a_n \xs_n - \x_n \sa _n \right) \right) 
\, \chi ( \bx , s ) ;
\label {W}
\eeq
both $ W( \ba , s ) $ and $ \chi ( \bx , s ) $
are even functions. 
It follows now from the expansion (\ref {GTrFEs})
that the s-ordered quasi-probability
distribution $ W( \ba , s ) $ 
is the weight function for the density \op $ \rho $
in the expansion
\beq
\rho = \int \! d^2\ba \, W( \ba , s ) \,  \tilde E( \ba , - s ) .
\label {rhoWE}
\eeq
\par
The functions $ W( \ba , s ) $ for different values
of the order parameter $ s $
are intimately related to one another
because the characteristic functions obey the identity
\beq
\chi( \bx , s ) = \exp \left( \frac { s }{ 2 }
\bxs \cdot \bx \right) \chi( \bx )
= \exp \left( \frac { ( s - t ) }{ 2 }
\bxs \cdot \bx \right) \chi( \bx , t ) .
\label {chichi}
\eeq
The function $ W( \ba , s ) $ is therefore the Fourier
transform of the product of $ \exp \left( \frac { ( s - t ) }{ 2 }
\bxs \cdot \bx \right) $ with the characteristic function
$ \chi( \bx , t ) $
\beq
W( \ba , s ) = \int \! d^2\bx \,
\exp \left( \sum_n
\left( \a_n \xs_n - \x_n \sa _n \right) \right)
\, \exp \left( \frac { ( s - t ) }{ 2 }
\bxs \cdot \bx \right) \, \chi( \bx , t ) .
\label {Wexxpchi}
\eeq
The Fourier transform of the characteristic function
$ \chi( \bx , t ) $
is $ W( \ba , t ) $, while that of
$ \exp \left( \frac { ( s - t ) }{ 2 }
\bxs \cdot \bx \right) $ according to Eq.(\ref {ex3}) is
\beq
\int \! d^2\bx \, e^ { \sum_n
\left( \c_n \xs_n - \x_n \cs _n \right) }
\, e^ { \frac { ( s - t ) }{ 2 }
\bxst \cdot \bxt } 
= \prod_n \left[ \frac { ( t - s ) }{ 2 } \, e^ { 
\frac { 2 } { ( t - s ) } \c_n \cs_n } \right] .
\label {convexp}
\eeq
The convolution theorem (\ref {convo}) now gives
$ W( \ba , s ) $ as
\beq
W( \ba , s ) =
\int \! \prod_j \left[ \frac { ( t - s ) }{ 2 } d^2\b_j \right] \,
\exp \left[
\frac { 2 } { ( t - s ) } \sum_i ( \a_i - \b_i ) ( \sa_i -\sb_i ) 
\right] W( \bb , t ) .
\label {Wst}
\eeq

\par
A useful example of $ W( \ba , s ) $ is the function 
\beq
W( \ba , -1 ) = \int \! d^2\bx \exp \left( \sum_n
\left( \a_n \xs_n - \x_n \sa _n \right) \right)
\, \chi ( \bx , -1 ) ,
\label {WA}
\eeq
which according to Eq.(\ref {chiaft}) is the Fourier transform
\beq
W( \ba , -1 ) = \int \! d^2\bx \, d^2\bb \,
\exp \left[ \sum_n
\left( \left( \a_n -\b_n \right) \xs_n 
- \x_n \left( \sa_n - \sb_n \right) \right) \right]
\, \langle \bb | \rho | - \bb \rangle .
\label {WA2}
\eeq
By using the delta-function identity (\ref {Delta}),
we see that this expression reduces to
\bea
W( \ba , -1 ) & = & \int \! d^2\bb \,
\d ( \ba - \bb ) 
\, \langle \bb | \rho | - \bb \rangle \nn\\
& = & \langle \ba | \rho | - \ba \rangle .
\label {WA3}
\eea
This function is the fermionic analogue
of the function $ Q( \bb ) = \langle \bb | \rho | \bb \rangle $
which is often used to represent the density operator $ \rho $
in terms of the bosonic coherent states $ | \bb \rangle $. 
It is the weight function that gives the mean values
of anti-normally ordered products of \cre and \ani \ops 
in terms of integrals of the corresponding products
of Grassmann numbers.

\section{Mean Values of Operators}
We shall here be concerned with computing
the mean values of the products of s-ordered monomials
\beq
\prod_i \,
\{ \left( a^\dgr_i \right)^{n_i} \, a_i^{ m_i } \}_s 
\label {prodmon}
\eeq
in which the exponents $ n_i $ and $ m_i $ take the values 0 or 1.
The ordering of the modes labelled by the index i is arbitrary 
but fixed.
We shall show that we may express
the mean values of such products of monomials
as integrals of the s-ordered weight function $ W( \ba , s ) $
multiplied by the monomials     in the same order.
By using the definition (\ref {W})
of $ W( \ba , s ) $,
we may write these integrals in the form
\bea
\leq{
\int \! d^2 \ba \, 
\prod_i \left( \sa_i \right) ^ { n_i } \a_i ^{ m_i } 
\, W( \ba , s ) } \nn\\
& = &
\int \! d^2 \ba \, 
\prod_i \left( \sa_i \right) ^ { n_i } \a_i ^{ m_i }
\, \int \! d^2\bx \exp \left( \sum_j
\left( \a_j \xs_j - \x_j \sa _j \right) \right) 
\, \chi ( \bx , s ) .
\label {step1}
\eea
It is now easy to write the monomial as 
a multiple derivative,
\bea
\leq{
\int \! d^2 \ba \,
\prod_i \left( \sa_i \right) ^ { n_i } \a_i ^{ m_i }
\, W( \ba , s ) } \nn\\
& = & 
\int \! d^2 \ba \, d^2\bx \,
\prod_i \left[ \frac { \p^{ n_i } } { \p ( - \x_i )^{ n_i } }
e^{ - \x_i \sa_i } \,
\frac { \p^{ m_i } } { \p ( - \xs_i ) ^{ m_i } }
e^{ - \xs_i \a_i } 
\right] \, 
\chi ( \bx , s ) .
\label {step2}
\eea
On using our formula (\ref {intbyparts})
for integration by parts, we have
\bea
\leq{
\int \! d^2 \ba \,
\prod_i \left( \sa_i \right) ^ { n_i } \a_i ^{ m_i }
\, W( \ba , s ) } \nn\\
& \!\!\! = \!\!\! &
\int \! d^2\bx \, d^2 \ba 
\exp \left( \sum_j \left( \a_j \xs_j - \x_j \sa _j \right) \right) 
\prod_i \left[
\frac { \p^{ n_i } } { \p ( \x_i )^{ n_i } } 
\frac { \p^{ m_i } } { \p ( \xs_i ) ^{ m_i } }
\right] \chi ( \bx , s ) 
\label {step3}
\eea
in which we recognize the delta-function formula (\ref {delta})
which gives
\bea
\leq{
\int \! d^2 \ba \,
\prod_i \left( \sa_i \right) ^ { n_i } \a_i ^{ m_i }
\, W( \ba , s ) } \nn\\
& = &
\int \! d^2\bx \, \d ( \bx ) \,
\prod_i 
\frac { \p^{ n_i } } { \p ( \x_i )^{ n_i } } \,
\frac { \p^{ m_i } } { \p ( \xs_i ) ^{ m_i } }
\, \chi ( \bx , s )
\label {step4} \\
& = &
\left. 
\prod_i
\frac { \p^{ n_i } } { \p ( \x_i )^{ n_i } } \,
\frac { \p^{ m_i } } { \p ( \xs_i ) ^{ m_i } }
\, \chi ( \bx , s ) \, \right|_{ \bx = 0 } \\
& = &
\left.
\Tr \left[
\rho \, \prod_i \,
\frac { \p^{ n_i } } { \p ( \x_i )^{ n_i } } \,
\frac { \p^{ m_i } } { \p ( \xs_i ) ^{ m_i } }
\left(
1 + \x_i a^\dgr_i + \xs_i a_i 
+ \xs_i \x_i \, \{ a^\dgr_i a_i \}_s
\right) \right|_{ \bx = 0 } \right] 
\quad 
\label {meanvals1}
\eea
If we recall the definitions of s-ordering
in Eqs.(\ref {adas}) and (\ref {trivdef}), we then find 
\beq
\int \! d^2 \ba \,
\prod_i \left( \sa_i \right) ^ { n_i } \a_i ^{ m_i }
\, W( \ba , s ) 
= \Tr \left[
\rho \, \prod_i \,
\{ \left( a^\dgr_i \right)^{n_i} \, a_i^{ m_i } \}_s \right] .
\label {meanvals}
\eeq
\par
In particular, by taking $ n_i = m_i = 0 $,
we see that the weight function $ W( \ba , s ) $
is normalized,
\beq
\int \! d^2 \ba \, W( \ba , s ) = \Tr \rho = 1 .
\label {Wnorm}
\eeq

\section{The P-Representation}
\par
Of the representations (\ref {rhoWE})
for the density \op $ \rho $, by far the most important
is the one for $ s = 1 $ with the normally ordered
weight function $ P( \ba ) = W( \ba , 1 ) $.
By Eq.(\ref {E-1}) it takes the simple form
\beq
\rho = \int \! d^2 \ba \, P( \ba ) \, | \ba \rangle \langle - \ba | ,
\label {Prep1} 
\eeq
which recalls the P representation~\cite{rjg2}--\cite{GDW}
for boson fields.
Since the function $ P( \ba ) $ is even,
we may also write
\beq
\rho = \int \! d^2 \ba \, P( \ba ) \, | - \ba \rangle \langle \ba | .
\label {Prep2}
\eeq
Because Grassmann integration is differentiation,
the fermionic P representation is not affected
by the mathematical limitations~\cite{rjg2}--\cite{kc2} 
that restricted somewhat the use of the bosonic
P representation.
\par
The P-representation
may be used directly to compute the mean values
of normally ordered products
\bea
\Tr \left( \rho \, a_k^{\dgr n} a^m_l \right)
& = & \int \! d^2 \ba P( \ba ) \langle \ba |
a_k^{\dgr n} a^m_l | \ba \rangle \nn\\
& = & \int \! d^2 \ba P( \ba ) \a_k^{\dgr n} \a_l^m .
\label {useofP}
\eea
This extremely useful relation 
is just a special case of Eq.(\ref {meanvals}) for $ s = 1 $\@.
\par
Since the \op $ \tilde E( \ba , 1 ) $ is
the dyadic (\ref {E+1}) of the \eiss of the
\cre \opsp, it follows from the expansion (\ref {rhoWE})
that the weight function (\ref {WA3})
\beq
Q( \ba ) \equiv W( \ba , -1 ) = \langle \ba | \rho | - \ba \rangle
\label {Qa}
\eeq
is the weight function in the representation
\beq
\rho = \int \! d^2 \ba \, Q( \ba ) \,
| \ba \rangle^\prime \, ^\backprime\langle - \ba | ,
\label {rhoQ}
\eeq
which affords the simple way of computing
the mean values of anti-normally ordered products
that corresponds to Eq.(\ref {meanvals}) for $ s = -1 $\@.
\par
Another use of the weight function
$ Q( \ba ) = W( \ba , -1 ) = \langle \ba | \rho | - \ba \rangle $,
however, is that it allows us to compute
the weight function $ P( \ba ) = W( \ba , 1 ) $
of the P-representation as the simple convolution
\beq
P( \ba ) = \int \! \prod_m \left( - d^2 \b_m \right) \,
\exp \left[ - \sum_n \left( \a_n - \b_n \right) 
\left( \sa_n - \sb_n \right) \right] \,
\langle \bb | \rho | - \bb \rangle ,
\label {PQ}
\eeq
as follows from the general convolution
formula (\ref {Wst}) with $ s = 1 $ and $ t = -1 $\@. 
Although the analogous relation for bosons 
often is singular~\cite{rjg2}--\cite{kc2},
this result holds for all fermionic density \ops $ \rho $.

\section{Correlation Functions for Fermions}
\par
A principal use of the P representation
for bosonic fields has been the evaluation
of the normally ordered correlation functions,
which play an important role in the theory
of coherence and of the statistics of photon-counting
experiments~\cite{rjg2}.
The analogously defined correlation functions
for fields of fermionic atoms can be shown to play 
a similar role in the description of atom-counting
experiments~\cite{rjg3}.
If we use $ \psi(x) $ to denote the positive-frequency part
of the Fermi field as a function of a space-time variable $x$,
then the first two of these correlation functions
may be defined as
\bea
G^{(1)} ( x , y ) & = & \Tr \left[ \rho \psi^\dgr (x) \psi(y) \right]
\label {G1}\\
G^{(2)} ( x_1 , x_2 , y_2 , y_1 ) & = & 
\Tr \left[ \rho \psi^\dgr (x_1) \psi^\dgr (x_2) 
\psi(y_2) \psi(y_1) \right] .
\label {G2}\\
\eea
The $n$th-order correlation function is
\beq
G^{(n)} ( x_1 , \dots, x_n , y_n , \dots, y_1 ) =  
\Tr \left[ \rho \psi^\dgr (x_1) \dots \psi^\dgr (x_n) 
\psi(y_n) \dots \psi(y_1) \right] .
\label {Gn}
\eeq
\par
If we expand the positive-frequency part
of the Fermi field in terms of its mode functions $ \phi_k( x ) $ as
\beq
\psi ( x ) = \sum_k a_k \phi_k( x ) ,
\label {psimodes}
\eeq
then its eigenvalue in the coherent state $ | \ba \rangle $
\beq
\psi ( x ) | \ba \rangle
= \varphi( x ) | \ba \rangle
\label {phieig}
\eeq
is the Grassmann field
\beq
\varphi ( x ) = \sum_k \a_k \phi_k( x ) 
\label {varphi}
\eeq
in which the \ani \ops in (\ref {psimodes})
are replaced by the Grassmann variables
$ \ba = \{ \a_k \}$.
\par
We may use the P representation to evaluate
the $n$th-order correlation function $ G^{(n)} $ 
as the integral 
\bea
\lefteqn{G^{(n)} ( x_1 , \dots, x_n , y_n , \dots, y_1 ) \qquad 
\qquad \qquad \qquad \qquad } \\
& = & \int \! d^2 \ba \, 
P( \ba ) \langle \ba |
\psi^\dgr (x_1) \dots \psi^\dgr (x_n)
\psi(y_n) \dots \psi(y_1) | \ba \rangle \\
& = & \int \! d^2 \ba \,
P( \ba ) \,
\varphi^\dgr (x_1) \dots \varphi^\dgr (x_n)
\varphi(y_n) \dots \varphi(y_1) .
\label {PG}
\eea

\section{Chaotic States of the Fermion Field}
The reduced density \op for a single mode
of the fermion field can be represented 
by a $ 2 \times 2 $ matrix for the states
with occupation numbers 0 and 1.
If the matrix is diagonal, it is specified completely
by the mean number of quanta $ \langle n \rangle $
in the mode.
The density \op for the $ k $th mode,
in other words, must take the form
\beq
\rho_k = \left( 1 - \langle n_k \rangle \right)
| 0 \rangle \langle 0 |
+ \langle n_k \rangle | 1 \rangle \langle 1 | .
\label {rhok}
\eeq
We shall speak of this density \op as representing
a chaotic state of the $ k $th mode.
A chaotic state of the entire field
will then be represented as a direct product
of such density \ops for all the modes of the field,
\beq
\rho_{ch} = \prod_k \rho_k .
\label {rhoch}
\eeq
It is specified by the complete set of mean 
occupation numbers $ \{ \langle n_k \rangle \} $.
\par
The total number of fermions,
$ N = \sum_k \ad_k a_k $,
present in chaotic states will in general be indefinite.
Indeed it is easily seen that in the state specified by
Eq.(\ref {rhoch}) we have
\beq
\langle N^2 \rangle - \langle N \rangle^2
= \sum_k \langle n_k \rangle ( 1 - \langle n_k \rangle )
\label {Nch}
\eeq
so that $ N $ can not be fixed unless all the 
$ \langle n_k \rangle $ take the values 0 or 1.
The indefiniteness of the number of particles present
is a feature that the chaotic states of the 
fermion and boson fields have in common.
For sufficiently large values of $ N $,
however, the fluctuations of $ N / \langle N \rangle $
may be quite small so the specification of $ N $
in these relative terms may be quite precise.
Fluctuations of this type in the number of particles present
are a familiar property of the grand canonical ensemble
in statistical mechanics,
and that ensemble, as we shall see, represents a special class
of chaotic states.
\par
The single-mode density \op (\ref {rhok}) 
can also be written as 
\beq
\rho_k = \left( 1 - \langle n_k \rangle \right)
\, \left( \frac { \langle n_k \rangle }{ 1 - \langle n_k \rangle }
 \right) ^{ \ad_k a_k } \,
\left( | 0 \rangle \langle 0 | 
+ | 1 \rangle \langle 1 | \right) 
\label {rad}
\eeq
in which we recognize the unit operator $ I_k $ for 
the subspace of the $ k $th mode.
Within this subspace we have
\beq
\rho_k = \left( 1 - \langle n_k \rangle \right)
\, \left( \frac { \langle n_k \rangle }{ 1 - \langle n_k \rangle }
 \right) ^{ \ad_k a_k } .
\label {rada}
\eeq
This expression can be used quite directly
to evaluate the weight function $ W( \ba , -1 ) = Q( \ba )$.
\par
We first note that for any real number $ v $
\beq
v^{ \ad a } | \a \rangle = 
e ^ { \shalf \a \sa ( 1 - v^2 ) } | \a v \rangle ,
\label {vid}
\eeq  
so that we have 
\bea
\langle \a | v^{ \ad a } | - \a \rangle & = &
e ^ { \shalf \a \sa ( 1 - v^2 ) } \,
\langle \a | - \a v \rangle \nn\\
& = & e ^ { \a \sa ( 1 + v ) } .
\label { vadid}
\eea
Then if we let $ v = \langle n_k \rangle / ( 1 - \langle n_k \rangle ) $,
we see that 
\beq
\langle \a_k | \rho_k | - \a_k \rangle = 
( 1 - \langle n_k \rangle ) \, \exp \left( \frac { \a_k \sa_k }
{ 1 - \langle n_k \rangle } \right) 
\label {arma}
\eeq
and
\bea
Q( \ba ) & = & W( \ba , -1 ) 
= \prod_k \langle \a_k | \rho_k | - \a_k \rangle \nn\\
& = & \prod_k ( 1 - \langle n_k \rangle ) 
\, \exp \left( \frac { \a_k \sa_k }
{ 1 - \langle n_k \rangle } \right) .
\label {Wmo}
\eea
This product is the weight function appropriate 
to averaging anti-normally ordered \op products
in chaotic states. 
\par
We may find the weight functions corresponding to all the
other ordering schemes by using
the convolution (\ref {Wst}) with $ t = -1 $
and carrying out the required integration
with sufficient attention to the implicit minus signs.
The result for the $ k $th mode is
\beq
W_k( \a_k , s ) = - \frac{ s + 2 \langle n_k \rangle - 1 } { 2 } \,
\exp \left( - \frac{ 2 \a_k \sa_k } { s + 2 \langle n_k \rangle - 1 }
\right) ,
\label {Wks}
\eeq
and the weight function for the multi-mode field
is simply the product 
\beq
W( \ba , s ) = \prod_k W_k( \a_k , s ) .
\label {Ws}
\eeq
\par
Thus the function $ W_k( \a_k , 0 ) $,
which is analogous to the Wigner function for boson fields,
is given by
\beq
W_k( \a_k , 0 ) = - \left( \langle n_k \rangle - \thalf \right) \,
\exp \left( - \frac{ \a_k \sa_k } { \langle n_k \rangle - \thalf } 
\right) ,
\label {WWch}
\eeq
and the function $ W_k( \a_k , 1 ) $,
which is the analogue of the function $ P_k( \a_k ) $
for boson fields, is
\beq
W_k( \a_k , 1 ) \equiv P_k( \a_k ) = - \langle n_k \rangle \,
\exp \left( { - \frac{ \a_k \sa_k } { \langle n_k \rangle } } \right) .
\label {Pkch}
\eeq
The latter result is a particularly useful one
since there are many physical contexts that call
for the averaging of normally ordered products
of \ani and \cre \opsp.
For chaotic fields one may calculate all such averages 
as Grassmann integrals by making use of the fermionic
P-representation with $ P( \ba ) $ given by Eq.( \ref {Pkch}). 
\par
The minus signs in front of the expressions (\ref {WWch}) and
(\ref {Pkch}) may be somewhat surprising since 
these functions are the fermionic
analogues of quasi-probability densities that are predominantly positive 
for boson fields.
It is worth pointing out, therefore,
that these signs result from our convention
that defines $ d^2\a $ as $ d\sa d\a $.
Had we chosen the differential instead to be $ d\a d\sa $,
the signs would have been positive. 
\par
For a chaotically excited boson field, the P-representation
expresses the \dop as a gaussian integral of a diagonal
\co_st dyadic.  For fermion fields the corresponding
expression of $ \rho_k $ for a single mode is
\beq
\rho_k = - \langle n_k \rangle \,
\int \! d^2 \a_k \, e^ { - \frac{ \a_k \sa_k }{ \langle n_k \rangle } }
\, | \a_k \rangle \langle - \a_k | .
\label {rhochPk}
\eeq
According to Eq.(\ref {rad}),
the \dop $ \rho_k $ can also be written as a sum
over the $m$-fermion states as 
\beq
\rho_k = ( 1 - \langle n_k \rangle ) \,
\sum_{m_k=0}^1 \left( \frac{ \langle n_k \rangle }{ 1 - \langle n_k \rangle }
\right) ^ { m_k } \, | m_k \rangle \langle m_k | .
\label {rhochmk}
\eeq
What we have shown, in effect, is that the two expressions
are identical and that statistical averages 
can be evaluated by means of gaussian integrations
for fermions as well as for bosons.
The multi-mode \dop is represented, of course,
by the product of the single-mode \dopsp,
$ \rho = \prod_k \rho_k $.
\par
Fields in thermal equilibrium with a suitable
particle reservoir represent particular examples
of the kind of chaotic excitation we have been describing.  
If it is appropriate to describe such fields
by means of the grand canonical ensemble,
then their overall \dop may be written as
\beq
\rho = \frac { 1 }{ \Xi( \b , \mu ) }
\, e^ { - \beta ( H - \mu N ) } ,
\label {GCE}
\eeq
where $ \beta = 1/ k_B T $, $ \mu $ is the chemical 
potential, $ H $ is the hamiltonian for the system,
$ N $ is the particle number, and the normalizing factor
$ \Xi( \b , \mu ) $ is the grand partition function.
For a field with dynamically independent mode functions
labeled by the index $ k $, we can write 
\beq
H = \sum_k \e_k \ad_k a_k , \qquad N = \sum_k \ad_k a_k
\label {HN}
\eeq
where $ \e_k $ is the energy of a particle in the $ k $th mode.
\par
Under these circumstances the equilibrium number 
of fermions in the $ k $th mode is 
\beq
\langle n_k \rangle = \frac{ 1 }{ e^ { \beta ( \e_k - \mu ) + 1 } } .
\label {Fermidist}
\eeq
In that case the ratio $ \langle n_k \rangle / ( 1 - \langle n_k \rangle ) $
is simply the generalized Boltzmann factor
\beq
\frac{  \langle n_k \rangle }{ 1 - \langle n_k \rangle }
= e^ { \beta ( \e_k - \mu ) } .
\label {BF}
\eeq
We then find that the product of the $ \rho_k $ given by Eq.(\ref {rhochPk})
is precisely equal to the grand canonical \dop (\ref {GCE}),
\beq
\int \! \prod_k \left( - \langle n_k \rangle d^2 \a_k 
e^ { - \frac{ \a_k \sa_k }{ \langle n_k \rangle } } \right) \,
| \ba \rangle \langle - \ba | = 
\frac { 1 }{ \Xi( \b , \mu ) }
\, e^ { - \beta ( H - \mu N ) } .
\label {climax}
\eeq
There are many examples of thermal equilibria
for which the P-representation on the left
should furnish a useful computational tool.

\section{Correlation Functions for Chaotic Field Excitations}
We have introduced a succession of normally ordered
correlation functions $ G^{(n)} ( x_1 \dots x_n , y_n \dots y_1 ) $
in section 13, and shown how they can be expressed 
as integrals over the Grassmann variables
$ \ba = \{ \a_k \} $.
For the case of chaotic fields, the appropriate weight function is 
\beq
P( \ba ) = \prod_k P_k( \a_k ) ,
\label {Pch}
\eeq
the product of the gaussian functions in Eq.(\ref {Pkch}).
The first-order correlation function is thus given by
\beq
G^{(1)} ( x , y ) = 
\int \! \prod_k \left( - \langle n_k \rangle d^2 \a_k
e^ { - \frac{ \a_k \sa_k }{ \langle n_k \rangle } } \right) \,
\langle \ba | \psi^\dgr( x ) \psi( y ) | \ba \rangle .
\label {G1ch}
\eeq
The fields $ \psi $ and $ \psi^\dgr $ may now be replaced
by their Grassmann field \eivs defined by (\ref {phieig}) 
and (\ref {varphi}).
Their product is a quadratic form
in the variables $ \a_k $ and $ \sa_k $,
which is easily integrated:
\bea
G^{(1)} ( x , y ) & = &
\int \! \prod_k \left( - \langle n_k \rangle d^2 \a_k
e^ { - \frac{ \a_k \sa_k }{ \langle n_k \rangle } } \right) \,
\sum_{l,m} \sa_l \a_m \phi_l^\dgr( x ) \phi_m( y ) \nn\\
& = & \sum_k \langle n_k \rangle \, \phi_k^\dgr( x ) \phi_k( y ) .
\label {G1phis}
\eea
\par
To find the higher-order correlation functions,
we can make use of a species of generating functional.
We first define the Grassmann fields
\bea
\zeta( x ) & = & \sum_k \b_k \phi_k( x ) \\
\eta( y ) & = & \sum_k \c_k \phi_k( y ) ,
\label {xeta}
\eea
and use them to construct the normally ordered
expectation value
\beq
\Gamma [ \zeta , \eta ] \equiv
\Tr \left[ \rho \exp \left( \int \! \zeta( x ) \psi^\dgr ( x ) d^4x \right)
\exp \left( \int \! \psi ( y ) \eta^*( y ) d^4y \right) \right] .
\label {Gamma}
\eeq
If we form the variational derivative of $ \Gamma $
with respect to $ \zeta( x_1 ) $ from the left 
and with respect to $ \eta^*( y_1 ) $ from the right,
subsequently setting $ \zeta $ and $ \eta $ to zero, 
then we find an alternative expression for the
first-order correlation function,
\beq
\frac{ \d }{ \d_L \zeta( x_1 ) }
\, \frac{ \d }{ \d_R \eta^*( y_1 ) } \left.
\Gamma \right|_{\zeta =\eta=0}
= \Tr \left[ \rho \psi^\dgr( x_1 ) \psi( y_1 ) \right]
= G^{(1)} ( x_1 , y_1 ) ,
\label {G1Gamma}
\eeq  
where left and right differentiation have been
indicated explicitly in the subscripts.
\par
It is evident then that one may generate
all of the higher-order correlation functions
by performing further differentiations, 
\beq
G^{(n)} ( x_1 \dots x_n , y_n \dots y_1 ) =
\frac{ \d }{ \d_L \zeta( x_1 ) } \dots
\frac{ \d }{ \d_L \zeta( x_n ) } \,
\frac{ \d }{ \d_R \eta^*( y_n ) } \dots
\frac{ \d }{ \d_R \eta^*( y_1 ) } \left. \Gamma \right|_{\zeta =\eta=0} .
\label {Gnzeta}
\eeq
To evaluate the generating functional $ \Gamma $
for a chaotic field, we make use of 
the orthonormality of the mode functions $ \phi_k $
and then carry out the Grassmann integration
\bea
\Gamma & = &
\int \! \prod_k \left( - \langle n_k \rangle d^2 \a_k
e^ { - \frac{ \a_k \sa_k }{ \langle n_k \rangle } } \right) \,
\exp \left( { \sum_l ( \b_l \sa_l + \a_l \cs_l ) } \right) \nn\\
& = & \prod_k \left( 1 + \langle n_k \rangle \b_k \cs_k \right) 
= \exp \left( { \sum_k \langle n_k \rangle \b_k \cs_k } \right) \nn\\
& = & \exp \left( { \int \! \zeta( x ) G^{(1)} ( x , y ) \eta^*( y ) 
d^4x d^4y } \right) .
\label {GammaG1}
\eea
If we begin performing the variational differentiations
to find the second-order correlation function,
we may write
\beq
\frac{ \d }{ \d_R \eta^*( y_2 ) } \,
\frac{ \d }{ \d_R \eta^*( y_1 ) } \left. \Gamma
\right|_{ \eta = 0 } = 
\int \! \zeta( x ) G^{(1)} ( x , y_2 ) d^4x \,
\int \! \zeta( x' ) G^{(1)} ( x', y_1 ) d^4x' .
\label {GaG1G1} 
\eeq
We then find 
\beq
G^{(2)} ( x_1 x_2 y_2 y_1 ) =  
\frac{ \d }{ \d_L \zeta( x_1 ) } 
\frac{ \d }{ \d_L \zeta( x_2 ) } 
\int \! \! \zeta( x ) G^{(1)} ( x , y_2 ) d^4x \!
\int \! \! \zeta( x' ) G^{(1)} ( x', y_1 ) d^4x' 
\eeq
and since $ \zeta( x ) $ and $ \zeta( x' ) $ anti-commute,
\beq
G^{(2)} ( x_1, x_2, y_2, y_1 )
=  G^{(1)} ( x_1, y_1 ) G^{(1)} ( x_2, y_2 ) 
- G^{(1)} ( x_1, y_2 ) G^{(1)} ( x_2, y_1 ) .
\label {G2G1G1}
\eeq 
\par
The generalization to $n$th order is immediate.
It expresses the $n$th-order correlation function
for chaotic fields as a sum of products of first-order 
correlation functions with permuted arguments,
\beq
G^{(n)} ( x_1 \dots x_n , y_n \dots y_1 ) =
\sum_P ( -1 )^P \prod_{j=1}^n G^{(1)} ( x_j , y_{Pj} ) .
\label {Gnperm}
\eeq 
This expression is summed over the $n!$ permutations
of the indices $ 1 \dots n $.
The factor $ ( -1 )^P $ is the parity of the permutation,
and the index $ Pj $ is the index that replaces $ j $
in the permutation.
\par
The expression of the $n$th-order correlation function
in terms of first-order correlation functions
is characteristic of chaotic fields.
Such fields are completely specified by the set
of mean occupation numbers $ \langle n_k \rangle $,
and these are already contained in the first-order 
correlation function.

\section{Fermion Counting Experiments}
The use of photon counting techniques has for many years 
been the most direct means of
investigating the statistical properties of light beams. 
Experiments of this type began with that of Hanbury Brown
and Twiss~\cite{HBT} in 1956 and expanded greatly in scope         
with the development of the laser. 
The theory~\cite{GDW} underlying these experiments is based 
on the evaluation of quantum-mechanical expectation values of
normally ordered products of electromagnetic field \opsp.
The \costs of the field~\cite{rjg2} thus play a special role 
in the formulation of that theory. 
The application of the theory, furthermore, extends 
to boson fields of much more general sorts, 
including for example beams of heavy atoms~\cite{rjg3}.
\par
In the case of the electromagnetic field,
it has been shown~\cite{GDW} that the probability of detecting 
$ n $ photons in a given interval of time can be expressed 
as the $n$th derivative with respect to a parameter $ \lambda $
of a certain generating function $ \cal{Q} ( \lambda ) $,
\beq
p( n ) = \frac { (-1)^n }{ n! } \frac { d^n }{ d \lambda^n }
\, \left. \cal{Q} ( \lambda ) \right|_{ \lambda = 1 } .
\label {pQ}
\eeq
The generating function $ \cal{Q} ( \lambda ) $
for the electromagnetic field is the expectation value of a normally
ordered exponential function of the form
\beq
\mathcal{Q} ( \lambda ) = 
\Tr \left( \rho \mbox{:} e^ { - \lambda \mathcal{ I } } \mbox{:} 
\right) ,
\label {Q}
\eeq
in which the symbols $ \mbox{:} \quad \mbox{:} $ stand for normal ordering, 
and the operator $ \cal{ I } $ is a space-time integral of the
product of the positive-frequency and negative-frequency parts
of the field, $ E^{ (+) }$ and $ E^{ (-) }$, respectively.
\par
For the case of fermion fields, 
it can easily be shown~\cite{rjg3} that the probability of counting 
$ n $ fermions in a given interval of time falls into precisely 
the same general form. 
In the simplest instance,
for detectors that respond to the density rather than the flux
of the particles, the integral $ \cal{ I } $ 
takes the form 
\beq
\mathcal{ I } = \kappa \, 
\int \! \psi^\dgr ( \vec r , t ) \psi ( \vec r , t ) 
d^3r dt ,
\label {Ik}
\eeq
where the constant $ \kappa $ is a measure of the 
sensitivity of the counter and the integration is
carried out over the counting-time interval and over 
the volume being observed.
\par
To obtain the expectation value of the 
normally ordered exponential function in Eq.(\ref {Q}),
then we may use the P-representation for the \dop $ \rho $.
In that case the field operators $ \psi ( \vec r , t ) $  
and $ \psi^\dgr ( \vec r , t ) $ are, in effect,
always applied to their \eissp, \costs such as $ | \ba \rangle $
and $ \langle \ba | $.  They can then be replaced
by their Grassmann field eigenvalue functions defined
by Eq.(\ref {varphi}) and its adjoint, so that
we have 
\beq
\mathcal{Q} ( \lambda ) = \int \! d^2 \ba P( \ba ) 
e ^ { - \lambda \mathcal{ J } } ,
\label {QJ}
\eeq
where 
\beq
\mathcal{ J } =  \kappa \, 
\int \! \varphi^* ( \vec r , t ) \varphi ( \vec r , t )
d^3r dt .
\label {Jk}
\eeq
The expression $ \mathcal{ J } $ is a quadratic form
that we can write as
\beq
\mathcal{ J } = \sum_{k, k'} \sa_k B_{ k k' } \a_{ k' } ,
\label { Jaa }
\eeq
so the evaluation of the generating function 
$ \mathcal{Q} ( \lambda ) $ reduces to the
calculation of the integral
\beq
\mathcal{ Q } ( \lambda ) = \int \! d^2 \ba P( \ba )
\, \exp \left( - \lambda \sum_{k, k'} \sa_k B_{ k k' } \a_{ k' }
\right) ,
\label {QB}
\eeq
in which the normal ordering symbols are no longer
necessary because of the simple anti-commutation
properties of the Grassmann variables $ \a_k $\@.
\par
For the case of the chaotic fields defined
in section 14,
this integral takes the form
\beq
\mathcal{ Q } ( \lambda ) = 
\int \! \prod_k \left( - \langle n_k \rangle d^2 \a_k
e^ { - \frac{ \a_k \sa_k }{ \langle n_k \rangle } } \right)
\, \exp \left( - \lambda \sum_{k, k'} \sa_k B_{ k k' } \a_{ k' }
\right) .
\label {Qch}
\eeq
If we define a new set of variables
$ \beta_k = \a_k / \sqrt{ \langle n_k \rangle } $,
we find according to the rule (\ref {ex1})
that the integral can be written as 
\beq
\mathcal{ Q } ( \lambda ) =
\int \! \prod_k ( - d^2 \b_k ) \,
\exp \left( \sum_{k, k'} \b^*_k
\left( \d_{ k k' } - \lambda M_{ k k' } \right) \b_{ k' }
\right) ,
\label {QBM}
\eeq
where the matrix $ M $ is
\beq
M_{ k k' } = \sqrt{ \langle n_k \rangle }
B_{ k k' } \sqrt{ \langle n_{ k' } \rangle } .
\label {MB}
\eeq
A unitary linear transformation
on the variables $ \b_k $ can then be used
to diagonalize the quadratic form in brackets.
If the \eivs of the matrix 
$ 1 - \lambda M $  are $ \mu_l $,
then the integral is easily seen,
according to the formula (\ref {ex3})
for $ \a = 0 $, to be
\beq
\mathcal{ Q } ( \lambda ) =
\prod_l \mu_l = \det \left( 1 - \lambda M \right) .
\label {Q1mM}
\eeq
This result may be used directly to find
the various probabilities given by Eq.(\ref {pQ}).
It contrasts quite interestingly
with the generating function
for boson counting distributions,
which with closely corresponding definitions
takes the form~\cite{GDW}
\beq
\mathcal{ Q }_B ( \lambda ) =
\frac { 1 }{ \det \left( 1 + \lambda M \right) } .
\label {Qb}
\eeq

\section{Some Examples}
\subsection{The Vacuum State}
For the density operator 
\beq
\rho = | 0 \dots 0 \rangle \langle 0 \dots 0 |
\label {vac}
\eeq
which represents the multi-mode vacuum state,
the normally ordered characteristic function
$ \chi_N( \bx ) $ is
\bea
\chi( \bx )_N & = & \Tr \left[ \rho
\, \exp\left( \sum_n \x_n a^\dgr_n \right)
\, \exp\left( - \sum_n a_n \xs_n \right)
\right] \nn\\
& = & \langle 0 \dots 0 | \exp\left( \sum_n \x_n a^\dgr_n \right)
\, \exp\left( - \sum_n a_n \xs_n \right) | 0 \dots 0 \rangle
= 1 . \quad
\label {vax}
\eea
The weight function of the P representation is then
\beq
P( \ba ) = \int \! d^2\bx \, \exp 
\left( \sum_i \left( \a_i \sx_i - \x_i \sa_i \right) \right)
= \d( \ba ) .
\label {pax}
\eeq
The mean values of the normally ordered products
of \cre and \ani \ops all vanish 
\beq
\Tr \left[
\rho \, \prod_i \,
\left( a^\dgr_i \right)^{n_i} \, a_i^{ m_i } \right]
= \int \! d^2 \ba \, \prod_i \,
\left( \sa_i \right)^{n_i} \, \a_i^{ m_i } \, 
\d( \ba ) = 0
\label {0}
\eeq
except for the trace
\beq
\Tr \left[ \rho  \right] =
\int \! d^2 \ba \, \d( \ba ) = 1 .
\label {Tr1}
\eeq
The general weight function $ W( \a , s) $
of the vacuum is given by 
\beq
W( \ba , s) = 
\half ( 1 - s ) \exp \left( \frac {2 \ba \cdot \bsa }{ (1 - s ) } \right) .
\eeq
 
\subsection{A Physical Two-Mode Density Operator}
Let us consider the most general physical two-mode
fermionic density operator
\bea
\rho & = & r \, | 0 0 \rangle \langle 0 0 |
+ u \, | 1 0 \rangle \langle 1 0 |
+ v \, | 0 1 \rangle \langle 0 1 |
+ w \, | 1 0 \rangle \langle 0 1 |
+ w^* \, | 0 1 \rangle \langle 1 0 | \nn\\
& & \mbox{} \qquad + x \, | 0 0 \rangle \langle 1 1 |
+ x^* \, | 1 1 \rangle \langle 0 0 |
+ t \, | 1 1 \rangle \langle 1 1 |
\label {rho1}
\eea
in which $ | 1 0 \rangle = a_1^\dgr | 0 0 \rangle $,
$ | 1 1 \rangle = a_2^\dgr  a_1^\dgr | 0 0 \rangle $, 
\emph {etc.}, and the Latin letters $ r, t, u, $ and $v $
represent non-negative real numbers, while $ x $ and 
$ w $ may be complex.  The non-zero traces are 
\bea
\Tr \rho & = & r + u + v + t = 1 \label {traces10} \\
\Tr \rho \, a_1^\dgr a_1 & = & u + t \\
\Tr \rho \, a_2^\dgr a_2 & = & v + t \\
\Tr \rho \, a_2^\dgr a_1 & = & w \\
\Tr \rho \, a_1^\dgr a_2 & = & w^* \\
\Tr \rho \, a_1 a_2 & = &  x^* \\
\Tr \rho \, a_1^\dgr a_2^\dgr & = & - x \\
\Tr \rho \, a_2^\dgr a_1^\dgr \, a_1 a_2 & = & t .
\label {traces11}
\eea
If the fermion number $ N $ commutes with the
density \op $ \rho $,
then $ x = x^* = 0 $.
\par
The normally ordered characteristic function
$ \chi_N( \bx ) $ is
\bea
\chi( \bx )_N \! & = & \! \Tr \left[ \rho 
\left( 1 + \x_1 a_1^\dgr - a_1 \xs_1 + \xs_1 \x_1 a_1^\dgr a_1 \right) 
\left( 1 + \x_2 a_2^\dgr - a_2 \xs_2 + \xs_2 \x_2 a_2^\dgr a_2 \right) 
\right] \nn\\
\! & = & \! 1 + w \xs_1 \x_2 + w^* \xs_2 \x_1 
+ ( u + t ) \xs_1 \x_1 + ( v + t ) \xs_2 \x_2 \nn\\
\! & & \! \mbox{} \qquad + x \, \x_1 \x_2 + x^* \, \sx_2 \sx_1
+ t \, \xs_1 \x_1 \xs_2 \x_2.
\label {chi1n}
\eea
\par
The weight function $ W( \ba , 1 ) $
is the Fourier transform of the normally ordered 
characteristic function $ \chi_N( \bx ) $
\bea
W( \ba , 1 ) \!\! & = & \!\! 
\int \! d^2\x_1 d^2\x_2 
\left( 1 + \a_1 \xs_1 + \sa_1 \x_1 + \sa_1 \a_1 \xs_1 \x_1 \right) \nn\\
& & \mbox{} \times
\left( 1 + \a_2 \xs_2 + \sa_2 \x_2 + \sa_2 \a_2 \xs_2 \x_2 \right) \nn\\
& & \mbox{} \times
\left[ 1 + w \, \xs_1 \x_2 + w^* \, \xs_2 \x_1
+ ( u + t ) \, \xs_1 \x_1 + ( v + t ) \, \xs_2 \x_2 \right. \nn\\
& & \qquad \left. \mbox{} + x \, \x_1 \x_2 + x^* \, \sx_2 \sx_1 
+ t \, \xs_1 \x_1 \xs_2 \x_2 \right], \nn
\eea
and after following the rules (\ref{I0}--\ref{intdef}), we find
\bea
W( \ba , 1 ) & = & 
t + w \a_2 \sa_1 + w^* \a_1 \sa_2 
+ ( v + t ) \sa_1 \a_1 + ( u + t ) \sa_2 \a_2 \nn\\
& & \qquad \mbox{} + x \, \a_1 \a_2 + x^* \, \sa_2 \sa_1
+ \sa_1 \a_1 \sa_2 \a_2.
\label {w12}
\eea
\par
We may now use this weight function to compute
the mean values
\bea
\int \! d^2\a_1 d^2\a_2 \, W ( \ba , 1 ) & = & \Tr \rho 
= 1 \label {traces12} \\
\int \! d^2\a_1 d^2\a_2 \, \sa_1 \a_1 \, W ( \ba , 1 ) & = & 
\Tr \rho \, a_1^\dgr a_1 = u + t \\ 
\int \! d^2\a_1 d^2\a_2 \, \sa_2 \a_2 \, W ( \ba , 1 ) & = & 
\Tr \rho \, a_2^\dgr a_2 = v + t \\ 
\int \! d^2\a_1 d^2\a_2 \, \sa_2 \a_1 \, W ( \ba , 1 ) & = & 
\Tr \rho \, a_2^\dgr a_1 = w \\
\int \! d^2\a_1 d^2\a_2 \, \sa_1 \a_2 \, W ( \ba , 1 ) & = & 
\Tr \rho \, a_1^\dgr a_2 = w^* \\
\int \! d^2\a_1 d^2\a_2 \, \a_1 \a_2 \, W ( \ba , 1 ) & = &
\Tr \rho \, a_1 a_2 =  x^* \\
\int \! d^2\a_1 d^2\a_2 \, \sa_1 \sa_2 \, W ( \ba , 1 ) & = &
\Tr \rho \, a_1^\dgr a_2^\dgr = - x \\
\int \! d^2\a_1 d^2\a_2 \, \sa_2 \sa_1 \a_1 \a_2 \, W ( \ba , 1 ) 
& = & \Tr \rho \, a_2^\dgr a_1^\dgr \, a_1 a_2 = t  
\label {traces13}
\eea
which agree with the results (\ref{traces10}--\ref{traces11}).
\par
With $ P( \ba ) = W ( \ba , 1 ) $ as given by (\ref {w12}),
we may write the density operator (\ref {rho1}) 
in the form of the fermionic P representation
\beq
\rho = \int \! d^2 \ba \, P( \ba ) \, | - \ba \rangle \langle \ba |
= \int \! d^2 \ba \, P( \ba ) \, | \ba \rangle \langle - \ba | .
\label {Prepex1}
\eeq


\begin{thebibliography}{99}
\bibitem{symcool} H.~C. Stoof, M.~Houbiers, C.~A. Sackett,
and R.~G. Hulet, {\sl Phys.~Rev. Letters\/} 76 (1996) 10;
C.~J. Myatt, E.~A. Burt, R.~W. Ghrist, E.~A. Cornell, and
C.~E. Wieman, {\sl Phys.~Rev. Letters\/} 78 (1997) 586.
\bibitem{rjg2} R.~J. Glauber, {\sl Phys.~Rev.\/ \rm 130} (1963) 2529; 
131 (1963) 2766.
\bibitem{GDW} R.~J. Glauber in
{\sl Quantum Optics and Electronics,}
eds.~C.~DeWitt {\sl et al.} (Gordon and Breach, 1965),
pp.~65--185 (especially pp.~178--182).
\bibitem{kcrg1} K.~Cahill and R.~J. Glauber,
{\sl Phys.~Rev.\/ \rm 177} (1969) 1857, 1882. 
\bibitem{kc2} K.~Cahill, {\sl Phys.~Rev.\/ \rm 180}
(1969) 1239, 1244. 
\bibitem{JS4} J.~Schwinger, {\sl  Phys.~Rev.\/} 92 (1953) 1283.
\bibitem{b} Grassmann algebra is conventionally used
in the formulation of the path integral for fermion fields.
It is discussed in a number of texts:
F.~A. Berezin, {\sl The Method
of Second Quantization\/}
(Academic Press, New York, 1966);
S.~Weinberg, {\sl The Quantum Theory of Fields:
Volume I: Foundations\/} (Cambridge University Press, 1995);
J.~Zinn-Justin, {\sl Quantum Field Theory
and Critical Phenomena\/} (Oxford University Press, 1989);
Lowell Brown, {\sl Quantum Field Theory\/}
(Cambridge University Press, 1992).
\bibitem{HBT} R.~Hanbury Brown and R.~Q. Twiss,
{\sl Nature\/} 177 (1956) 27.
\bibitem{rjg3} R.~J. Glauber, Heineman Prize Lecture,
D.A.M.O.P.~Meeting of the A.P.S., 
Ann Arbor, MI, May 1996.
\end{thebibliography}
\end{document}